%% file: mugdha.tex
\documentclass[12pt]{article}

\input preamble.tex

\title{CNN vs ELM for Image-Based Malware Classification}

\author{Mugdha Jain\footnotemark[1]\ \ \ 
William Andreopoulos\footnotemark[1]\ \ \
Mark Stamp\footnotemark[1]\,\,\footnotemark[2]}

\begin{document}

\symbolfootnotetext[1]{Department of Computer Science, San Jose State University}
\symbolfootnotetext[2]{mark.stamp$@$sjsu.edu}

\maketitle

\abstract
Research in the field of malware classification often relies on
machine learning models that are trained on high-level features, 
such as opcodes, function calls, and control flow graphs. 
Extracting such features is costly, since disassembly or 
code execution is generally required. In this paper, we conduct 
experiments to train and evaluate machine learning models for malware classification,
based on features that can be obtained without disassembly or execution of code. 
Specifically, we visualize malware samples as images and employ image analysis
techniques. In this context, we focus on two machine learning models, namely,
Convolutional Neural Networks (CNN) and Extreme Learning Machines (ELM).
Surprisingly, we find that ELMs can achieve accuracies on par with CNNs, 
yet ELM training requires less than~2\%\  of the time
needed to train a comparable CNN.

\section{Introduction}\label{chap:intro}

Malware is software that is designed to disrupt or damage computer systems. 
According to Symantec, more than~$669$ million new 
malware variants were detected in~2017, which was an increase of more
than~80\% from~2016~\cite{symantec}. Clearly, malware detection is a critical 
task in computer security.

Commonly used malware detection techniques can be broadly categorized into anomaly-based 
and signature-based. However, these strategies have some potential disadvantages. 
Obfuscation can be used to evade 
signature-based detection~\cite{Xu2013}, while anomaly-based detection is costly and 
often yields an unacceptably high false positive rate~\cite{falsepos}. Malware detection 
based on machine learning models may overcome these weaknesses. 

Successful machine learning based malware detection approaches have been
trained on high-level features such as opcode sequences, function calls, or control flow 
graphs~\cite{Farrokhmanesh2019, Hashemi2017, Santos:2013:OSR:2442161.2442234}. 
However, extracting such high-level features can be costly; hence approaches that do not
require extensive pre-processing are preferred, provided that sufficient accuracy can be attained. 
Specifically, we prefer features that can be obtained without disassembly or code execution.

In this research, we consider the Malimg dataset~\cite{Nataraj:2011:MIV:2016904.2016908},
which consists of malware samples that have been visualized as images.
To obtain these malware images, bytes from executable files are 
trivially mapped to pixels to create grayscale 
images~\cite{Nataraj:2011:MIV:2016904.2016908}. The primary focus of our research is to
compare the classification performance of convolutional neural networks (CNN) 
and extreme learning machines (ELM) for these malware images. 

CNNs have been shown to perform well in image classification tasks~\cite{cnn}.
In comparison, ELMs are a somewhat less well-known technique, 
particularly in the field of malware analysis. 
ELMs are somewhat controversial, but experimental studies have 
shown that they can produce acceptable predictive 
performance in some tasks, and at a much lower training
cost, as compared to networks that are trained by backpropagation~\cite{youYou}. 
This makes ELMs attractive for malware classification, particularly when considering the large 
volume of new malware variants. 

When testing CNN and ELM models on the Malimg dataset~\cite{Nataraj:2011:MIV:2016904.2016908},
we perform extensive parameter tuning to optimize the performance of each model. For ELMs we also 
consider an ensemble approach, as previous work has shown that such ensembles can overcome 
some of the instability issues that are inherent in ELMs~\cite{ensemble}.

The remainder of this paper is organized as follows. Section~\ref{chap:background} discusses
relevant previous work. In Section~\ref{chap:implementation} we provide a summary of the 
dataset used in this research, and we briefly discuss the machine learning techniques that are 
used. In Section~\ref{chap:results}, we present and analyze the results of our experiments. 
Finally, in Section~\ref{chap:conclusion}, we summarize our results and consider possible
avenues for future work.

\section{Background}\label{chap:background}

Due to the large volume of new malware, antivirus developers must 
constantly upgrade their methods and algorithms. This makes accurate and rapid 
detection of malware a critical topic in information security~\cite{Hashemi2017}.

Malware classification can be based on static or dynamic features, or some combination thereof. 
Static features are those that are extracted from static files, 
while dynamic features are extracted during code execution or emulation. 
Static approaches often use features such as calls to external libraries, strings, 
and byte sequences for classification~\cite{kolter}. Other static approaches 
extract higher-level information from binaries, such as sequences of API calls~\cite{cesare} or opcode information~\cite{Wong2006}. Examples of dynamic features include resource usage 
and the frequency of calls to specific kernel functions.

A considerable amount of previous work has been done on malware classification. 
In this section, we discuss a few representative samples of malware classification 
techniques that have appeared in literature.

\subsection{Previous Work}

Machine learning models for malware detection and classification are 
trained on features or attributes extracted from executable files.
As previously mentioned,
examples of such features include %byte $n$-grams,
opcodes, API calls, control flow graphs, and many others. In many cases
the extraction of these features can be costly, so approaches using raw bytes 
are preferred, if comparable accuracy can be obtained.
For example, byte~$n$-grams have been successfully used as features. 
As another example, it is possible to treat executable files
as images, and apply image analysis techniques.
%This latter approach is the focus of this paper.

%%%%% It gets a bit tedious to read when every paragraph starts "In~\cite{...},"
An opcode-based approach for malware classification is 
presented in~\cite{Santos:2013:OSR:2442161.2442234}. The opcodes are obtained by 
disassembling the corresponding \texttt{exe} files. 
The authors then use a technique for determining 
the relevance of an opcode that is similar to the well-known tf-idf 
(Term Frequency-Inverse Document Frequency) forumla, which is 
a numerical statistic often used in text mining and search engines. 
In this context,
opcode $n$-grams are considered as the ``words.'' 
These authors also define a Weighted Term Frequency (WTF) 
to measure of the relevance of each opcode. Although they obtain 
satisfactory results, their method has a number of disadvantages,
including the costly step of disassembling executable files to extract opcodes.

%%%%% Any advantage to this over PCA?????
Another opcode-based malware detection technique is proposed in~\cite{Hashemi2017}. 
This work relies on an adjacency matrix based on pairs of consecutive opcodes. 
The feature vector for each file is constructed by applying the ``power iteration'' 
method to the opcode graph. They obtain reasonable results using this technique for 
the problem of distinguishing malware from benign. However, the authors observe that 
the power iteration method generally takes a significant amount of time to converge 
to the dominant eigenvector.

In~\cite{924286}, the authors divide each executable into non-overlapping byte substrings
and the frequency histogram of these substrings comprise the feature vectors. 
They developed a data-mining method that is tested on parts of the file, 
such as the header or text (code) section. The authors claim that their 
technique improves detection rates for new malware, as compared to 
traditional signature-based methods.

The method in~\cite{kolter} is also based on byte $n$-grams. 
In this case, each $n$-gram is considered as a boolean attribute, which is \texttt{true} 
if the $n$-gram is present in an executable file and \texttt{false} otherwise. The top~500 
$n$-grams are selected based on Information Gain (IG). The authors test a variety
of machine learning algorithms on these features (SVM, Na\"{i}ve Bayes, and 
Decision Tree) and obtain good results for detecting unknown malware instances.

The method proposed in~\cite{santos2009n} relies on $n$-grams as file signatures. 
The authors employ the $k$-nearest neighbor algorithm and experiment with 
different values of $n$ for the $n$-grams. They achieve their best results 
using 4-grams.

In~\cite{7529495}, the authors propose a technique based on converting executable files to audio. 
Specifically, they convert each byte of an executable into its corresponding 
decimal value, which in turn represents a specific musical note in the MIDI 
encoding scheme. Using this technique, each executable is converted into a MIDI file.
%which contains a sequence of messages that encode the byte data into music. 
The MIDI file is then converted into an audio signal, and Music Information Retrieval 
(MIR) techniques are used to extract audio features. Their method achieves an accuracy 
in excess of~90\% for distinguishing malware from benign,
based on $k$-nearest neighbor, AdaBoost, and Random Forest classifiers.

Extreme Learning Machines (ELM) have been applied to malware detection on 
the Android platform in~\cite{ZhangAndroid}. In this work, ELMs are trained on 
static features, and the results are reasonably strong. In~\cite{ShahabELM}, 
the authors consider the effectiveness of High Performance Extreme Learning Machine (HP-ELM) 
by varying the features and activation functions of the HP-ELM. Their experiments yield 
a maximum accuracy of 95.92\%.

A two layer ELM is applied to the malware detection problem in~\cite{telm}. 
A partially connected network is used between the input and the first hidden layer. 
This layer is then aggregated with a fully connected network in the second layer.
The authors utilize an ensemble to improve the accuracy and robustness 
of the system.

In~\cite{Nataraj:2011:MIV:2016904.2016908}, the authors propose a method to directly 
map bytes from an executable file to pixels to create a grayscale image. From the resulting 
images, they observe that malware files belonging to the same family appear similar in 
layout and texture. They achieve a classification accuracy of 98\% 
using~$k$-nearest neighbor. The features considered are based on
so-called GIST descriptors, which summarize gradient information in an image. 
One clear advantage of this method is that neither disassembly nor 
code execution is required to obtain the features.

The research we present in this paper was motivated to some extent by image-based
analysis in~\cite{Nataraj:2011:MIV:2016904.2016908}, as well as the ELM-based
thread in~\cite{telm,ShahabELM,ZhangAndroid}.
In this paper, we focus on
the relative advantages and disadvantages of Convolutional Neural Networks (CNN) 
and Extreme Learning Machines (ELM) when trained directly on pixel data, 
rather than, say, high-level GIST descriptors. The features we consider are trivial
to extract. In addition, the ELM technique is extremely efficient to train,
and we obtain surprisingly strong results, as compared to CNNs.

Next, we provide details on relevant implementation and background topics. Then
in Section~\ref{chap:results} we present and discuss our experimental results.

\section{Implementation}\label{chap:implementation}

In this section, we give a summary of the malware families and the 
dataset used in this research. Also, we briefly discuss the machine learning 
techniques that we use in our experiments.

\subsection{Dataset}

We consider malware classification based on the Malimg 
dataset~\cite{Nataraj:2011:MIV:2016904.2016908}. 
This dataset contains in excess of~9300 grayscale images 
belonging to~25 different malware families. This dataset has been used
as the basis for several previous research papers, including~\cite{Sravani18}
and the 
aforecited~\cite{telm,Nataraj:2011:MIV:2016904.2016908,ShahabELM,ZhangAndroid},
among others.

%%%%% ?????
%%%%% Would be good to have a table of comparing our results to previous work that uses Malimg dataset
%%%%% ?????

Table~\ref{tab:families} lists the~25 families in the Malimg dataset
and the number of samples that belong to each family. Note that the Malimg dataset is 
extremely imbalanced. For example, the largest family, \texttt{Allaple.A}, has  nearly
one-third of all of the samples and it is more than~36 times as large as the
least numerous family, \texttt{Skintrim.N}, while the two largest families
comprise nearly half of the entire dataset.
This imbalance creates some
issues when attempting to classifying the samples.
We return to this imbalance issue in 
Section~\ref{sect:weightedELM}.

\begin{table}[!htb]
\begin{center}
\caption{Type of each malware family}\label{tab:families}
\begin{tabular}{lc|lc}\midrule\midrule
\textbf{Family} & \textbf{Samples} & \textbf{Family} & \textbf{Samples}\\
\midrule
\texttt{Adialer.C}		&\z125 	& \texttt{Lolyda.AA2}		&\z184 \\
\texttt{Agent.FYI}		&\z116 	& \texttt{Lolyda.AA3}		&\z123 \\
\texttt{Allaple.L}			&1591 	& \texttt{Lolyda.AT}		&\z159 \\
\texttt{Allaple.A}		&2949 	& \texttt{Malex.gen!J}	&\z136 \\
\texttt{Alueron.gen!J}		&\z198 	& \texttt{Obfuscator.AD}	&\z142 \\
\texttt{Autorun.K}		&\z106 	& \texttt{Rbot!gen}		&\z158 \\
\texttt{C2Lop.P}			&\z146 	& \texttt{Skintrim.N}		&\z\z80 \\
\texttt{C2Lop.gen!G}		&\z200 	& \texttt{Swizzor.gen!E}	&\z128 \\
\texttt{Dialplatform.B}	&\z177 	& \texttt{Swizzor.gen!I}	&\z132 \\
\texttt{Dontovo.A}		&\z162 	& \texttt{VB.AT}			&\z408 \\
\texttt{Fakerean}		&\z381 	& \texttt{Wintrim.BX}		&\z\z97 \\
\texttt{Instantaccess}		&\z431 	& \texttt{Yuner.A}		&\z800 \\ \cline{3-4}
\texttt{Lolyda.AA1}		&\z213 	& Total				& 9342\\
\midrule\midrule 
\end{tabular}
\end{center}
\end{table}

\subsubsection{Visualization of Malware Files}
In~\cite{Nataraj:2011:MIV:2016904.2016908}, the authors describe the technique used 
to convert malware executables into grayscale images. A given malware binary is read 
as a vector of~8 bit unsigned integers and then organized into a 2-dimensional 
array. This is visualized 
as a grayscale image where each pixel is in the range~0 to~255,
where~0 is black and~255 is white. The width of the image is fixed and the height 
is allowed to vary depending on the file size, as summarized in Table~\ref{tab:imagesize}.

\begin{table}[!htb]
\begin{center}
\caption{Image dimensions based on malware size}\label{tab:imagesize}
\begin{tabular}{c|c}\midrule\midrule
\textbf{File Size} & \textbf{Width (pixels)}\\
\midrule
less than 10kB          &\z\z32\\
10kB -- 30kB	      &\z\z64\\
30kB -- 60kB	         &\z128\\
60kB -- 100kB	        &\z256\\
100kB -- 200kB	        &\z384\\
200kB -- 500kB        &\z512\\
500kB -- 1000kB        &\z768\\
greater than 1000kB	         &1024\\
\midrule\midrule 
\end{tabular}
\end{center}
\end{table}

\begin{figure}[!htb]       
\centering
    \includegraphics[scale=0.55]{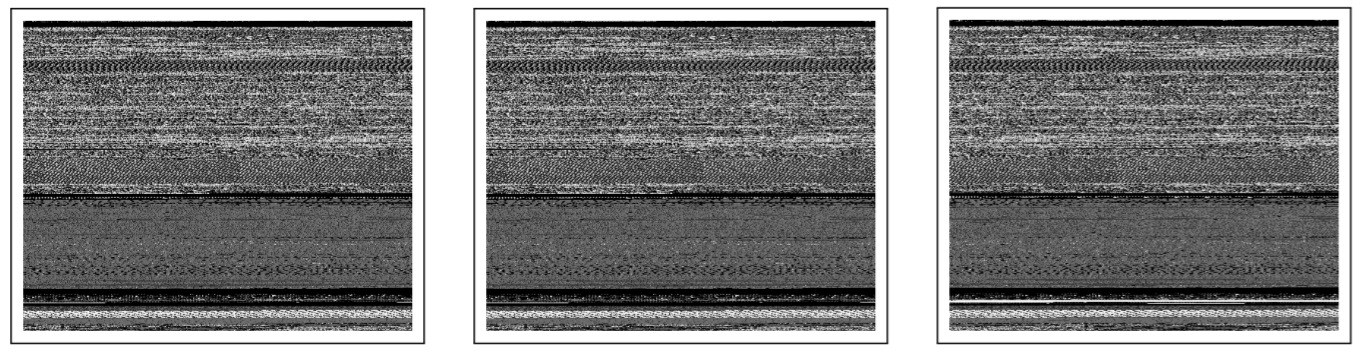}
    \caption{Adialer.C samples}\label{fig:adialercsamples}
\end{figure}

\begin{figure}[!htb]     
\centering
    \includegraphics[scale=0.55]{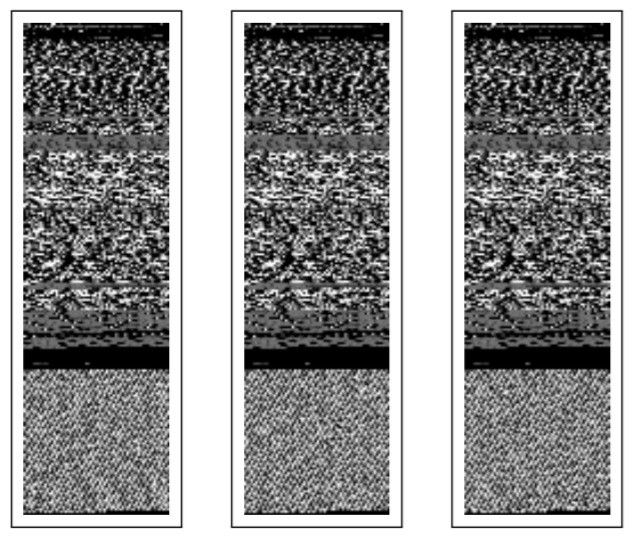}
    \caption{Dialplatform.B samples}\label{fig:dialplatformsamples}
\end{figure}

\begin{figure}[!htb]       
\centering
    \includegraphics[scale=0.55]{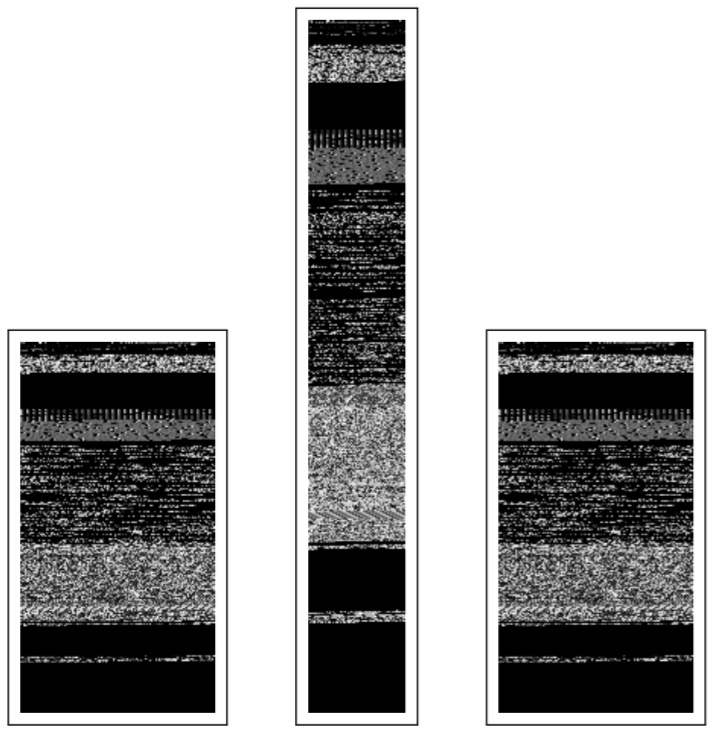}
    \caption{Dontovo.A samples}\label{fig:dontovosamples}
\end{figure}

Figures~\ref{fig:adialercsamples} through~\ref{fig:dontovosamples} show
show three different samples from each of three malware families. 
Specifically, Figure~\ref{fig:adialercsamples} has samples from the
\texttt{Adialer.C} family, while Figure~\ref{fig:dialplatformsamples} shows 
samples of \texttt{Dialplatform.B} family, and Figure~\ref{fig:dontovosamples} 
contains samples of the \texttt{Dontovo.A} family. 
These images indicate that there is considerable 
intra-family similarity and inter-family differences,
at least with respect to these three specific malware families. 
Intuitively, we would expect image-based analysis to perform
well on such data.

%\begin{figure}[!htb]       
%\centering
%    \fbox{\includegraphics[scale=0.2]{images/adialerc1.png}}   
%    \hspace{5px}
%    \fbox{\includegraphics[scale=0.2]{images/adialerc2.png}}
%    \hspace{5px}
%    \fbox{\includegraphics[scale=0.2]{images/adialerc3.png}}
%    \caption{Adialer.C samples}\label{fig:adialercsamples}
%\end{figure}

%\begin{figure}[!htb]     
%\centering
%    \fbox{\includegraphics[scale=0.6]{images/dialplatform1.png}} 
%    \hspace{5px}  
%    \fbox{\includegraphics[scale=0.6]{images/dialplatform2.png}}
%    \hspace{5px}
%    \fbox{\includegraphics[scale=0.6]{images/dialplatform3.png}}
%    \caption{Dialplatform.B samples}\label{fig:dialplatformsamples}
%\end{figure}

%\begin{figure}[!htb]       
%\centering
%    \fbox{\includegraphics[scale=0.4]{images/dontovo1.png}}   
%    \hspace{10px}
%    \fbox{\includegraphics[scale=0.4]{images/dontovo2.png}}
%    \hspace{10px}
%    \fbox{\includegraphics[scale=0.4]{images/dontovo3.png}}
%    \caption{Dontovo.A samples}\label{fig:dontovosamples}
%\end{figure}

\subsection{Classification Techniques}

In this section, we describe the machine learning models considered in
this paper. First, we discuss Convolutional Neural Networks and 
then we outline the main concepts behind Extreme Learning Machines.

\subsubsection{Convolutional Neural Networks}

Convolutional Neural Networks (CNN) draw inspiration from the workings and hierarchical 
structure of the primary visual pathway of the brain. In the~1950s and~1960s, 
experiments were conducted to understand how the brain perceives the 
world visually~\cite{hubel1962receptive}. It was discovered
that there are two basic types of visual neuron cells, namely, 
simple cells (S cells) and complex cells (C cells). These cells activate 
when they identify basic shapes such as lines in a fixed area and at a specific angle. 

A CNN is a type of feed-forward neural network for processing data 
that takes a 2-dimensional or higher-dimensional matrix as input. CNNs have
proven highly effective for image analysis. CNNs typically 
include multiple convolutional layers, one or more pooling layers, 
and a fully-connected output layer.
All of the CNN experimental results discussed in this 
paper are based on models trained using Keras~\cite{chollet2015keras}. 

%\begin{figure}[!htb]
%\centering
%\includegraphics[width=0.9\textwidth]{images/cnn.png}
%\caption{CNN Architecture~\cite{imagecnn}}\label{fig:cnn}
%\end{figure}

%\subsubsubsection{Convolutional Layer}

Convolutional layers are the main building blocks of CNNs.
A discrete convolution is a sequence that is itself 
a composition of two sequence---more precisely, 
a discrete convolution is
computed as a sum of pointwise products. Let~$c=x*y$ 
denote the convolution of sequences~$x=(x_0,x_1.x_2,\ldots)$ 
and~$y=(y_0,y_1.y_2,\ldots)$. Then~$c_k$, the~$k^{\thth}$
element of the convolution, is given by
$$
%  (x*y)_{k} = \sum_{k=i+j} x_{i} y_{j} = \sum_{i=-\infty}^{\infty} x_{i} y_{k-i} 
  c_{k} = \sum_{k=i+j} x_{i} y_{j} = \sum_{i} x_{i} y_{k-i} 
$$
We can view this process as~$x$ acting as a ``filter'' (or kernel) on the 
sequence~$y$ over a sliding window.
The concept of a discrete convolution 
is easily extended to higher-dimensional data.
With CNNs, we generally treat color images as 3-dimensional arrays,
with the third dimension given by the red, green, and blue color planes
of the RGB encoding.

In a CNN, the coefficients of the convolutions (i.e., weights)
are learned via training, with this training typically completed
via the backpropagation algorithm. Convolutions serve to greatly reduce the 
number of weights, as compared to a fully connected
network, and thus make training on images practical.
In addition, when applying CNNs to images, we obtain a high degree 
of translation invariance, which is highly desirable in image
analysis, and serves to greatly reduces the overfitting that
would otherwise tend to occur.

\begin{figure}[!htb]
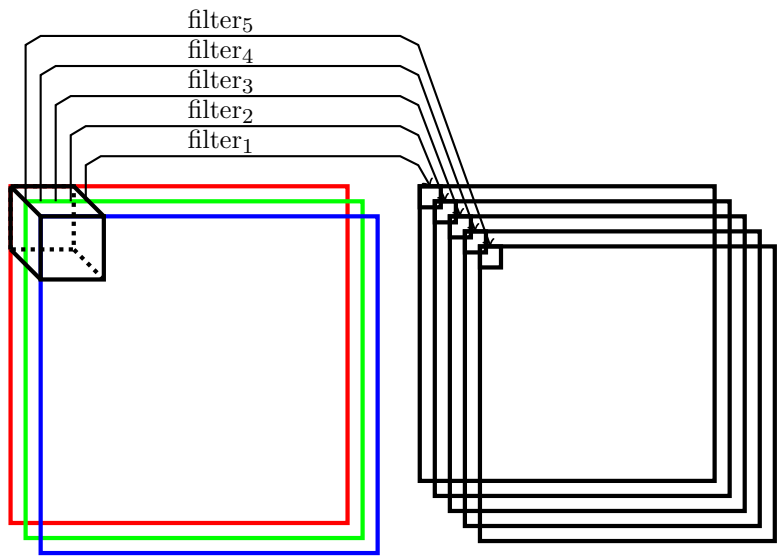

  \centering
    \input figures/conv3b.tex
  \caption{First convolutional layer with stack of~5 filters (RGB image)}\label{fig:convlayer}
\end{figure}

The first convolutional layer of a CNN is applied to the input image. 
This is illustrated in Figure~\ref{fig:convlayer}, where 
the input consists of an RGB image, which is treated as 3-dimensional
data. In this case, five filters are learned at the first layer---by randomly
initializing the weights of the filters, the CNN can learn
multiple filters at each layer. Thus, the next convolutional layer
can also be applied to higher-dimensional data.

After the first convolutional layer, subsequent convolutional layers
are applied to the output of previous convolutional layers. Hence,
we compute convolutions of convolutions.
At the first convolutional layer, the filters learn intuitive 
features, such as edges and basic shapes, whereas the second convolutional
layer learns more abstract features, such as ``texture.''
Convolutional layers learn progressively more abstract
features that ultimately can be used to distinguish, say,
a picture of a cat from that of a dog.

%Stride is the size of the step the convolution filter moves each time. 
%A stride size of 1 means that the filter slides pixel by pixel. 
%By increasing the stride size, the filter slides over the input with a 
%larger interval and thus has less overlap between the cells.

%\subsubsubsection{Pooling Layer}

\begin{figure}[!htb]
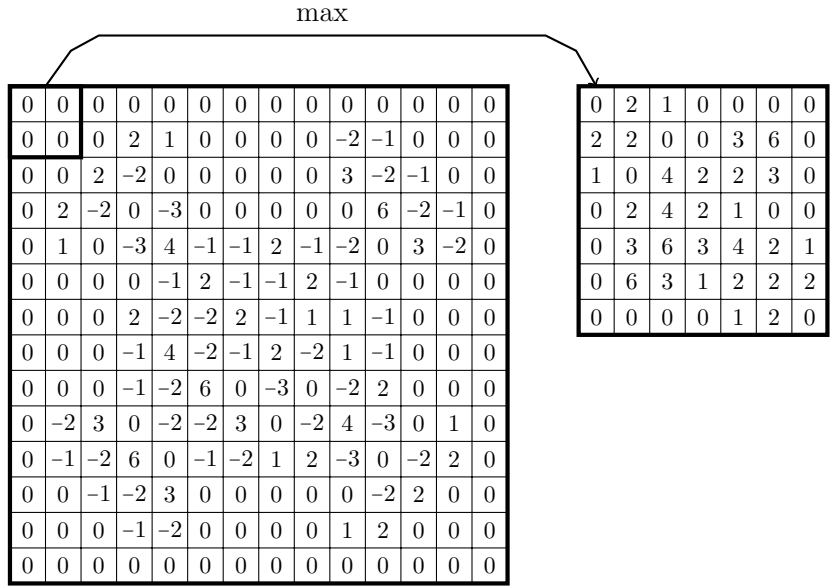

  \centering
    \input figures/maxPool.tex
  \caption{Max pooling layer ($2\times 2$)}\label{fig:maxpool}
\end{figure}

Between each pair of convolutional layers, it is common to add a pooling layer. 
The function of pooling is to reduce the dimensionality,
and thus reduce the training time. Pooling 
might also improve various desirable properties,
such as translation invariance. 
There are several types of pooling
functions, with max pooling being the most common.
Max pooling consists of simply selecting the maximum value over
a given filter block. Figure~\ref{fig:maxpool} illustrates max pooling 
over a 2-dimensional feature vector. 

%\begin{figure}[!htb]
%\centering
%\includegraphics[width=0.6\textwidth]{images/maxpool.png}
%\caption{Max Pooling~\cite{maxpool_layer}}\label{fig:maxpool}
%\end{figure}

Note that convolutional layers
typically use a stride of one, while pooling uses a stride 
equal to the filter width, which serves to further increase the 
downsampling effect of pooling. Recently, it has
become popular to use convolutional layers
with a stride greater than one in place of pooling layers.

\subsubsection{Extreme Learning Machine}\label{sect:ELMs}

As with most aspects of ELMs, the origin of the technique
is somewhat controversial. The unfortunate terminology of 
``Extreme Learning Machine'' was apparently first used 
in~\cite{elmoriginal}. Regardless of the origin of the technique,
ELMs are essentially randomized feedforward neural networks that 
effectively minimize the cost of training.

An ELM consists of a single layer of hidden nodes, 
where the weights between inputs and hidden nodes are 
randomly initialized and remain unchanged throughout training. 
The weights that connect the hidden nodes to 
the output are trained, but due to the simple structure of an ELM,
these weights can be determined by solving linear equations---more
precisely, by solving a linear regression problem. Since
no backpropagation is required, ELMs are far more efficient to train,
as compared to other neural network architectures. However, since
the weights in the hidden layer are not optimized, we will typically
require more weights in an ELM, which implies that the testing phase
may be somewhat more costly, as compared to a network
trained by backpropagation. Nevertheless, in applications 
where models must be trained frequently, ELMs
can be competitive.

%\begin{figure}[!htb]
%\centering
%\includegraphics[width=0.5\textwidth]{images/elmarch.png}
%\caption{Architecture of ELM Model~\cite{telm}}\label{fig:elmarch}
%\end{figure}

\begin{figure}[!htb]
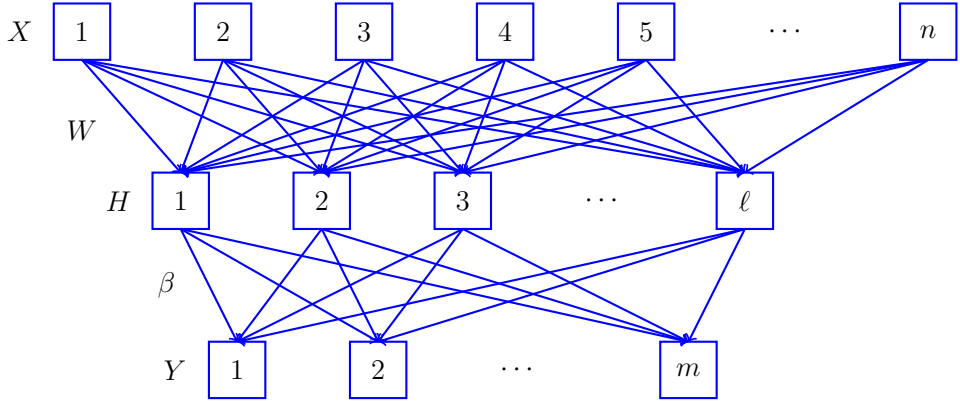

  \centering
    \input figures/elm.tex
\caption{Architecture of an ELM model}\label{fig:elmarch}
\end{figure}

Consider the ELM architecture shown in Figure~\ref{fig:elmarch}, 
where~$X$ denotes the input layer, $H$ is the hidden layer and~$Y$ is 
the output layer. In this example, there are~$N$ samples of the form~$(x_i, y_i)$
for~$i=1,2, \dots, N$, where $x_i=\rowvecccc{x_{i_1}}{x_{i_2}}{\dots}{x_{i_n}}^T$ 
is the feature vector for sample~$i$
and~$y_i = \rowvecccc{y_{i_1}}{y_{i_2}}{\dots}{y_{i_m}}^T$ are the output labels,
where~$T$ indicates the transposition operation.
Then the input and output for the ELM are 
as $X = \rowvecccc{x_1}{x_2}{\dots}{x_n}^T$ 
and $Y = \rowvecccc{y_1}{y_2}{\dots}{y_m}^T$, respectively.
In this example, the hidden layer~$H$ has~$\ell$ neurons. 
We denote the activation function of the hidden layer as~$g(x)$.

%%%%% Displayed equations should read as part of a sentence (as corrected for first one below)
%%%%% And, only number equations that you need to refer to somewhere else in the paper
%%%%% you can use "$$" for equations that do not need to be numbered
To train an ELM, we randomly selects the weight 
matrix that connects the input layer~$X$ to the 
hidden layer~$H$. We denote this randomly-assigned weight matrix 
as~$W = \rowvecccc{w_1}{w_2}{\dots}{w_{\ell}}$
where each~$w_i$ is a column vector.
We also randomly select the bias 
matrix~$B = \rowvecccc{b_1}{b_2}{\dots}{b_{\ell}}$ 
for this same layer. 
During the training phase, both~$W$ and~$B$ remain unchanged. 

After $W$ and $B$ have been initialized, 
the output of the hidden layer~$H$ is given by
$$
    H=g(WX + B) .
$$
The output of the ELM is denoted as~$Y$ and is calculated as
$$
  Y=H\beta
$$
where~$\beta$ is the weight matrix for the output layer.

The values of the weights~$\beta$ at the hidden layer 
are learned via linear least squares,
and can be computed using~$H^\dagger$, 
the Moore-Penrose generalized inverse of~$H$, as discussed below.
It is worth emphasizing that the only parameters that are learned in
the ELM are the elements of~$\beta$. 

Given that~$Y$ is the desired output, a unique solution of the 
system based on least squared error can be found as follows. 
We denote the Moore-Penrose generalization inverse of $H$ as~$H^\dagger$, 
which is defined as
$$
H^\dagger = \begin{cases}			
		(H^TH)^{-1}H^T  \text{ if $H^TH$ is nonsingular}\\
		H^T(HH^T)^{-1}  \text{ if $HH^T$ is nonsingular}
		\end{cases}
$$
Then the desired solution~$\beta$ is give by 
%%%%% Define any symbol before you use it
$$
\beta=H^\dagger Y
$$
%%%%% Not a new paragraph, so don't insert a blank line. In any case, don't start
%%%%% a sentence with a symbol, number, erc.

After calculating~$\beta$, the training phase ends. 
For each test sample~$x$, the output~$Y$ can be calculated as 
$$
  Y=g\bigl(C(x)\bigr) \beta
$$
where~$C(x)$ is defined below.
The entire training process is extremely efficient, particularly in comparison
to the backpropagation technique that is typically used to train neural 
networks~\cite{StampANN}.

For the research reported in this paper, 
we use the Python implementation of 
ELMs given in~\cite{python-elm}. This implementation 
uses input activations 
that are a weighted combination of two functions
that are referred to as 
an ``MLP'' kernel and an ``RBF'' kernel---we employ
the same terminology here. The MLP kernel
is simply the linear operation
$$
  \mlpKernel (x) = Wx + B
$$
where the weights~$W$ and biases~$B$ 
are randomly selected from a normal distribution.
This is the kernel function that is typically associated with 
a standard ELM.

The RBF kernel is considerably more complex, and is based 
on generalized radial basis functions as defined in~\cite{melm-grbf}.
%Note that in~\cite{melm-grbf}, this RBF kernel is applied to ELM training.
The details of this RBF kernel go beyond the scope of this paper;
see~\cite{melm-grbf} for additional information and, in particular,
examples where this kernel is applied to train ELMs.
We use the notation~$\rbfKernel(x)$ to represent
the RBF kernel. Also, it is worth noting that the RBF kernel 
is much more costly to compute, and hence its use does somewhat negate
one of the major advantages of an ELM. We provide timing comparisons for these
kernels in Section~\ref{chap:results}.

The input activations are given by
\begin{equation}\label{eq:inputAct}
  C(x) = \alpha \mlpKernel (x) + (1 - \alpha) \rbfKernel (x)
\end{equation}
where~$0\leq\alpha\leq 1$ is a user-specified mixing parameter.
Note that for~$\alpha=0$ we use only the MLP kernel~$\mlpKernel(x)$
and for~$\alpha=1$, only the RBF kernel~$\rbfKernel(x)$ is used.

\section{Experiments and Results}\label{chap:results}

In this section, we discuss the results of malware classification 
experiments involving the Malimg dataset.
First we consider our CNN experiments, 
then we discuss our ELM results. Finally, we compare the
results obtained using these two approaches.

\subsection{CNN Experiments}

For the experiments discussed in this section, 
each CNN model was trained for~50 epochs with the~\texttt{relu} 
activation function. Various CNN architectures with different combinations of 
convolution, pooling, and dense (fully-connected) layers are considered.
We also give results related to tuning of 
the hyperparameter, including input image size, batch size, number of filters,
filter size, and pooling size.

\subsubsection{One Convolutional Layer}\label{ssec:1Cmodels}

Our one-convolutional layer models were trained on 
different input image sizes ($32\times 32$, $64\times 64$, 
and $128\times 128$ pixels) and different number of 
filters ($32$  and~$64$). Each of these CNN models uses~$3\times 3$
filters, a fully connected layer with $128$ neurons, 
and an output layer with $25$~neurons corresponding to the classes 
in the Malimg dataset. A \texttt{softmax} activation function is used in each case. 
Table~\ref{tab:1cnn_result} summarizes our results
for these six cases.
We see that the larger image size of~$128\times 128$ clearly outperforms
the smaller image sizes.

%%%%% Be consistent in everything, such as number of decimal points of precision
%%%%% And usually better to decimal than percentages, especially since you use decimal elsewhere
%%%%% And don't put "px"
\begin{table}[!htb]
\begin{center}
\caption{CNN models with one convolutional layer}\label{tab:1cnn_result}
\begin{tabular}{c|c|c}\midrule\midrule
\textbf{Image Size} & \textbf{Filters}  &\textbf{Accuracy}\\
\midrule
$32\times 32$             &32		 & 0.8400\\
$32\times 32$            &64		 & 0.8467\\
$64\times 64$            &32		 & 0.9340\\
$64\times 64$            &64		 & 0.9245\\
$128\times 128$            &32		&  {\bftab 0.9630}\\
$128\times 128$            &64		 & 0.9589\\
\midrule\midrule 
\end{tabular}
\end{center}
\end{table}

\begin{figure}[!htb]
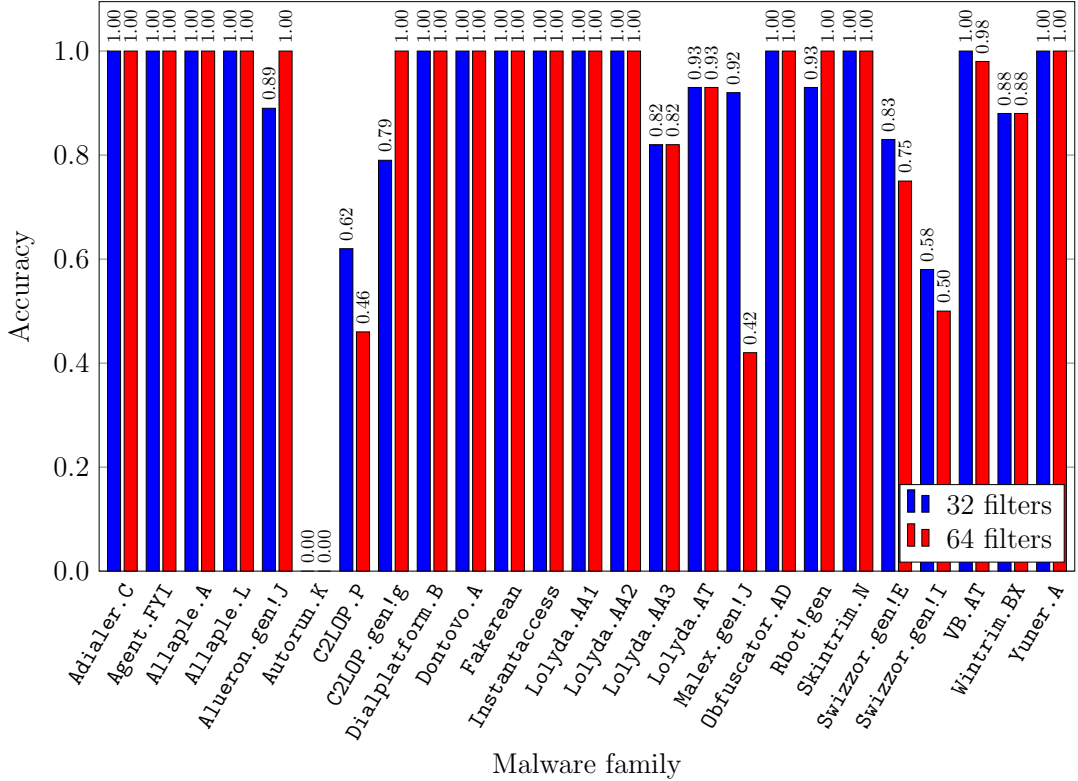

\centering
\input figures/fig_cnn_c1.tex
\caption{CNN with one convolutional layer ($128\times 128$ images)}\label{fig:cnn128_1c}
\end{figure}

\begin{figure}[!htb]
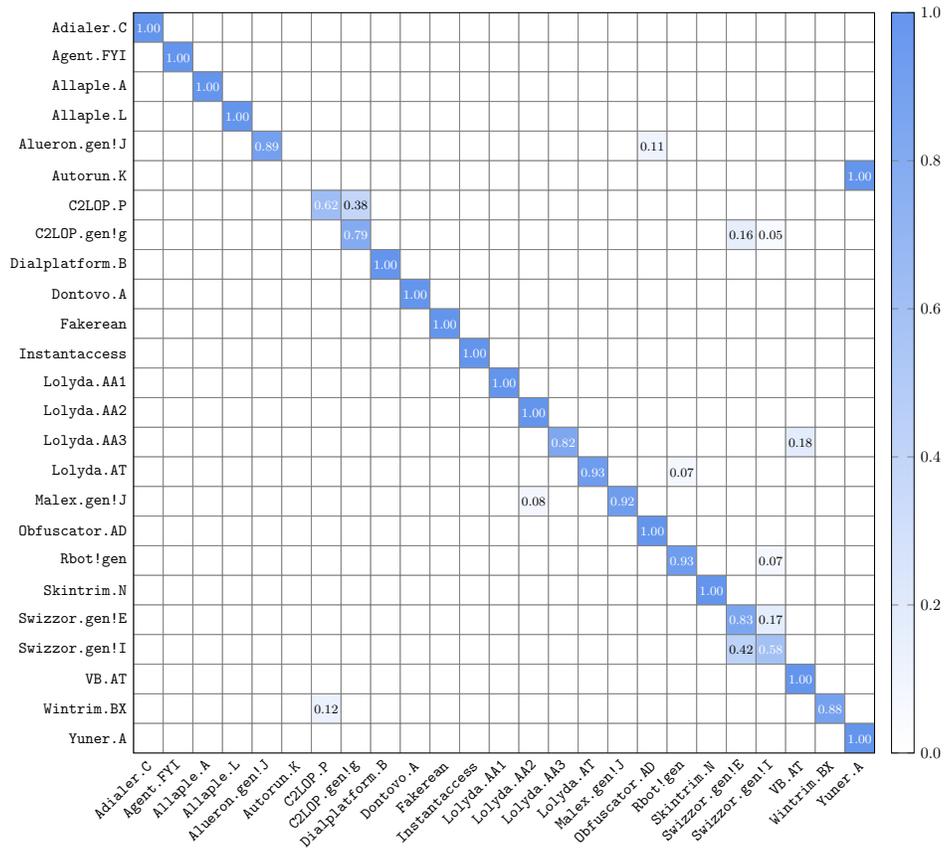

	\centering
	\input figures/conf_exp2.tex
\caption{Confusion matrix for best CNN model ($128\times 128$ images and~$32$ filters)} 
\label{fig:best_cm}
\end{figure}

Classwise accuracies for the~$128\times 128$
experiments in Table~\ref{tab:1cnn_result} 
are summarized in Figure~\ref{fig:cnn128_1c}.
We observe that
training these CNNs on images of size $128\times 128$ with $32$~filters 
results in an accuracy of $96.3\%$. 
%This is the best overall accuracy
%that we have obtained using CNN models in our research. 
The confusion matrix for this best case
is given in Figure~\ref{fig:best_cm}. This model is able to achieve an accuracy of 
above~90\%\ for~16 out of the~25 classes in the dataset---only one family, 
namely, \texttt{Autorun.K}, has an accuracy below~50\%. 

%\begin{figure}[!htb]
%\centering
%\begin{tabular}{c}
%\includegraphics[width=0.6\textwidth]{images/cnn128_1c_32f.png}
%\\
%(a) Image size $128\times 128$ with~$32$ filters
%\\
%\\
%\includegraphics[width=0.6\textwidth]{images/cnn128_1c_64f.png}
%\\
%(b) Image size $128\times 128$ with~$64$ filters
%\end{tabular}
%\caption{CNN with one convolutional layer}\label{fig:cnn128_1c}
%\end{figure}

%%%%% Adialer.C missing from original image ?????

\begin{figure}[!htb]       
\centering
    \begin{tabular}{ccc}
    \includegraphics[scale=0.5]{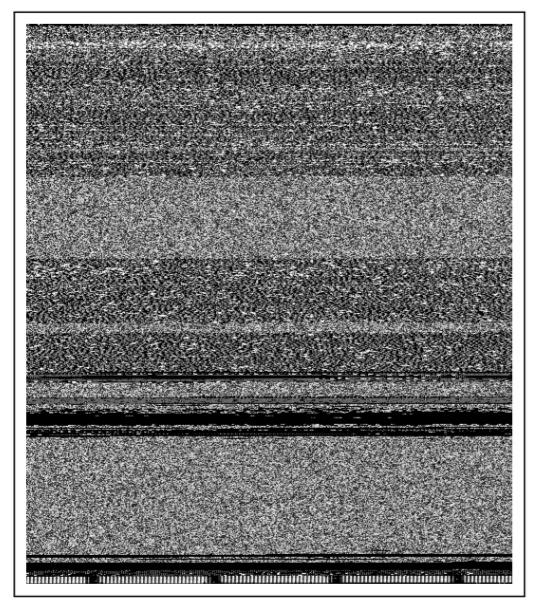}
    &
    &
    \includegraphics[scale=0.5]{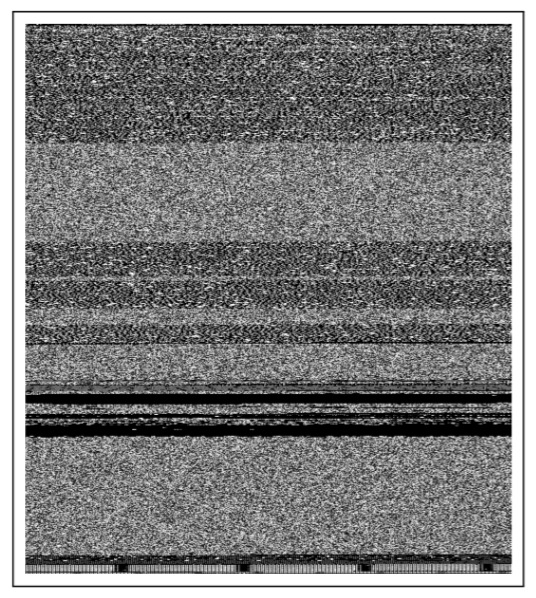}
    \\
    \\[-1ex]
    (a) \texttt{Swizzor.gen!I}
    &
    &
    (b) \texttt{Swizzor.gen!E}
    \end{tabular}
    \caption{Samples from \texttt{Swizzor} families}\label{fig:swiz_similar}
\end{figure}

The confusion matrix in Figure~\ref{fig:best_cm} also shows which specific
malware families are frequently misclassified. For example, we see that~$5$ 
of the~$12$ samples of the family \texttt{Swizzor.gen!I} are 
misclassified as belonging to \texttt{Swizzor.gen!E}, which is not surprising,
given that they are both variants of the Swizzor trojan family. The visual similarity 
between malware images from these families can be seen in 
Figure~\ref{fig:swiz_similar}. 

%\begin{figure}[!htb]       
%\centering
%    \begin{tabular}{ccc}
%    \fbox{\includegraphics[scale=0.25]{images/swizzorgeni1.png}}   
%    &
%    &
%    \fbox{\includegraphics[scale=0.25]{images/swizzorgene1.png}}
%    \\
%    \\[-1ex]
%    (a) \texttt{Swizzor.gen!I}
%    &
%    &
%    (b) \texttt{Swizzor.gen!E}
%    \end{tabular}
%    \caption{Samples from \texttt{Swizzor} families}\label{fig:swiz_similar}
%\end{figure}

\begin{figure}[!htb]       
\centering
    \begin{tabular}{ccc}
    \includegraphics[scale=0.5]{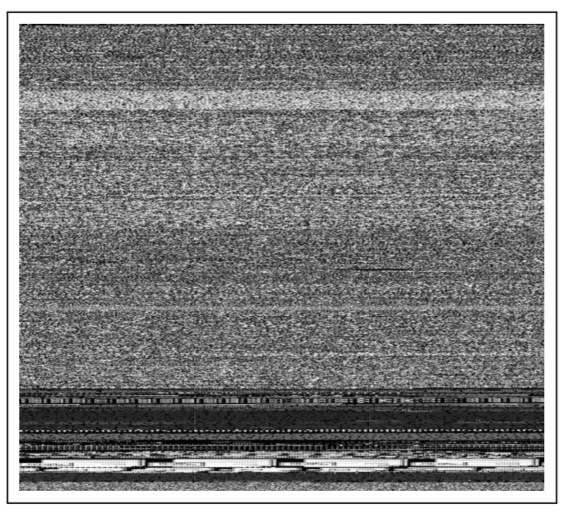}
    &
    &
    \includegraphics[scale=0.5]{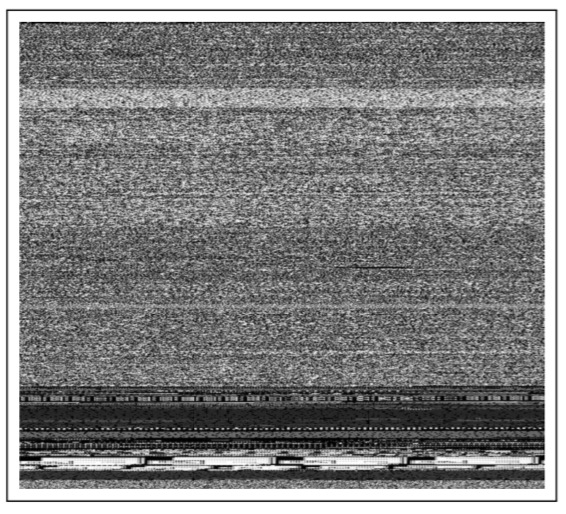}
    \\
    \\[-1ex]
    (a) \texttt{Autorun.K}
    &
    &
    (b)  \texttt{Yuner.A}
    \end{tabular}
    \caption{Samples from a pair of similar families}\label{fig:yuner_similar}
\end{figure}

Another interesting misclassification case occurs for samples 
of \texttt{Autorun.K}, which are all misclassified
as belonging to the family \texttt{Yuner.A}. The visual similarity between these 
two families is also quite obvious, as can be seen in Figure~\ref{fig:yuner_similar}.

%\begin{figure}[!htb]       
%\centering
%    \begin{tabular}{ccc}
%    \fbox{\includegraphics[scale=0.18]{images/autorunk1.png}}   
%    &
%    &
%    \fbox{\includegraphics[scale=0.18]{images/yunera1.png}}
%    \\
%    \\[-1ex]
%    (a) \texttt{Autorun.K}
%    &
%    &
%    (b)  \texttt{Yuner.A}
%    \end{tabular}
%    \caption{Samples from a pair of similar families}\label{fig:yuner_similar}
%\end{figure}

We note that for all three image sizes considered, increasing the number of filters has 
minimal effect on the one-convolutional layer CNN accuracy. This could be attributed 
to the fact that images in the dataset are in grayscale and have relatively
simple structure, and hence only a small number of filters is required.
In contrast, there is a definite increase in the overall 
accuracy when the size of the input 
image is increased. However, there are some anomalies,
such as \texttt{Autorun.K}, which is predicted with $100\%$ 
accuracy for all models based on image size~$32\times 32$, 
but for~$64\times 64$ and~$128\times 128$ size images,
samples of this family are consistently misclassified as \texttt{Yuner.A}.

\subsubsection{Two Convolutional Layers}\label{ssec:2Cmodels}

This section briefly describes the experiments conducted with 
CNN models with two convolutional layers. These models were trained 
for similar cases of image size, filters, pooling layer, etc., 
as in the previous section. However, since we have two convolutional
layers, we experiment with different numbers of filters in these layers.
Table~\ref{tab:2cnn_result} summarizes our results
for these two convolutional layer models.

\begin{table}[!htb]
\begin{center}
\caption{CNN models with two convolutional layers}\label{tab:2cnn_result}
\begin{tabular}{c|c|c|c}\midrule\midrule
\multirow{2}{*}{\textbf{Image Size}}  & \multicolumn{2}{c|}{\textbf{Number of Filters}} &
\multirow{2}{*}{\textbf{Accuracy}} \\
 & \textbf{First Layer}  &\textbf{Second Layer} \\
\midrule
$32\times 32$             &32	&32	 & 0.820\\
$32\times 32$             &32	&64	 & 0.816\\
$32\times 32$            &64	&32	 & 0.823\\
$32\times 32$            &64	&64	 & 0.834\\
$64\times 64$             &32	&32	 & 0.934\\
$64\times 64$             &32	&64	 & 0.928\\
$64\times 64$             &64	&32	 & 0.950\\
$64\times 64$             &64	&64	 & 0.943\\
$128\times 128$         &32	&32	 & 0.921\\
$128\times 128$         &32	&64	 &  {\bftab 0.957}\\
%%%%% Where are the other two 128/128 cases ?????
\midrule\midrule 
\end{tabular}
\end{center}
\end{table}

\begin{figure}[!htb]
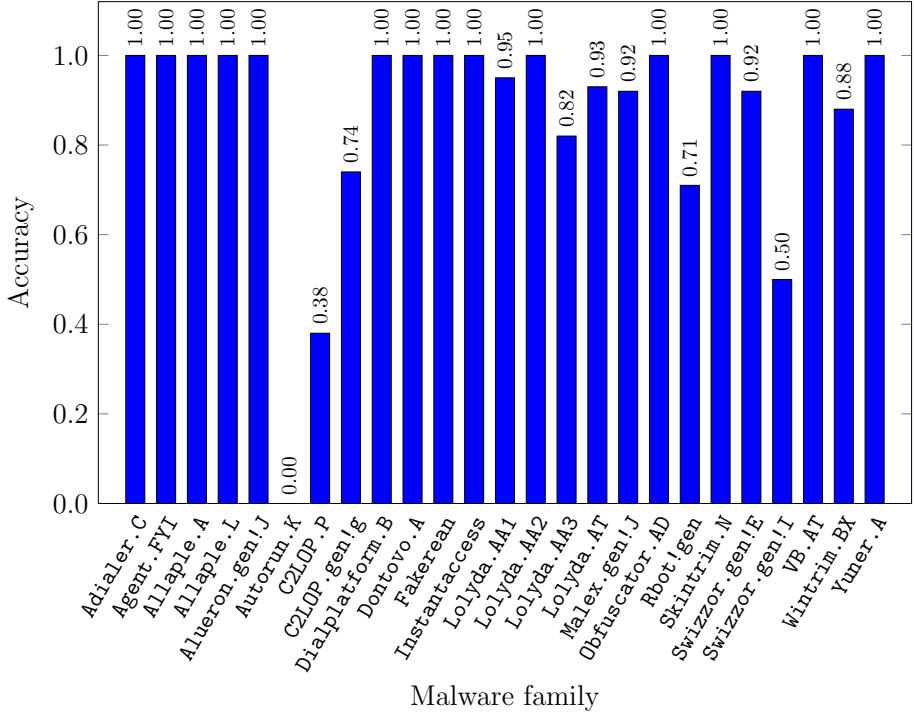

\centering
\input figures/fig_cnn_c2.tex
\caption{CNN with two convolutional layers, 
$128\times 128$ images, and~$(32,64)$ filters}\label{fig:2cnn128_32f_64f}
\end{figure}

The best results in Table~\ref{tab:2cnn_result} were obtained with
input images of size~$128\times 128$ pixels 
with~$32$ and~$64$ filter maps. This model achieved an overall accuracy 
of $95.67\%$.
The classwise accuracies for this case
are summarized in Figure~\ref{fig:2cnn128_32f_64f}.

%\begin{figure}[!htb]
%\centering
%\includegraphics[width=0.6\textwidth]{images/2cnn128_32f_64f.png}
%\caption{CNN with two convolutional layers 
%($128\times 128$ images and~$(32,64)$ filters)}\label{fig:2cnn128_32f_64f}
%\end{figure}

%%%%% Adialer.C missing from original image ?????

%The two convolutional layer models did not show improvement over 
%one convolutional layer modes discussed above. 
Our best two-convolutional layer model in Table~\ref{tab:2cnn_result}
is just slightly below the best one-layer model in Table~\ref{tab:1cnn_result}.
Since training time is significantly increased for two-layer models, 
and since a one-convolutional layer model performs best,
the one-layer model is the clearly the best among the CNN
architectures that we tested.

\subsection{ELM Experiments}

This section describes out ELM experiments and results. 
Analogous to the CNN experiments discussed in the previous section, 
we test a wide variety of ELM architectures. For ELMs, the number of parameters
are relatively few---we experiment with the number of neurons in the hidden layer 
and the choice of activation functions. 
For each combination of parameters, 
we train and test~50 ELMs, 
and evaluate the performance in terms of the average accuracy.
In addition, also consider ensembles
of ELMs, we perform experiments to analyze the stability of our models, 
and we consider the effect of regularization---in the form of 
dropouts. 

\subsubsection{Input Activation}\label{ssec:alpha1}

In this section, we experiment with the input activation function
parameter~$\alpha$, which appears in equation~\eref{eq:inputAct}.
Recall that the input activation function 
is a linear combination of a so-called MLP kernel
and an RBF kernel, with~$\alpha$ determining the relative weighting
of these two kernels.

We trained~$50$ ELMs on images of size~$64\times 64$, with $\alpha=1$.
For this choice of the parameter~$\alpha$, only the highly-efficient
MLP kernel is used. We refer to this as pure MLP
activation.
 
\begin{figure}[!htb]
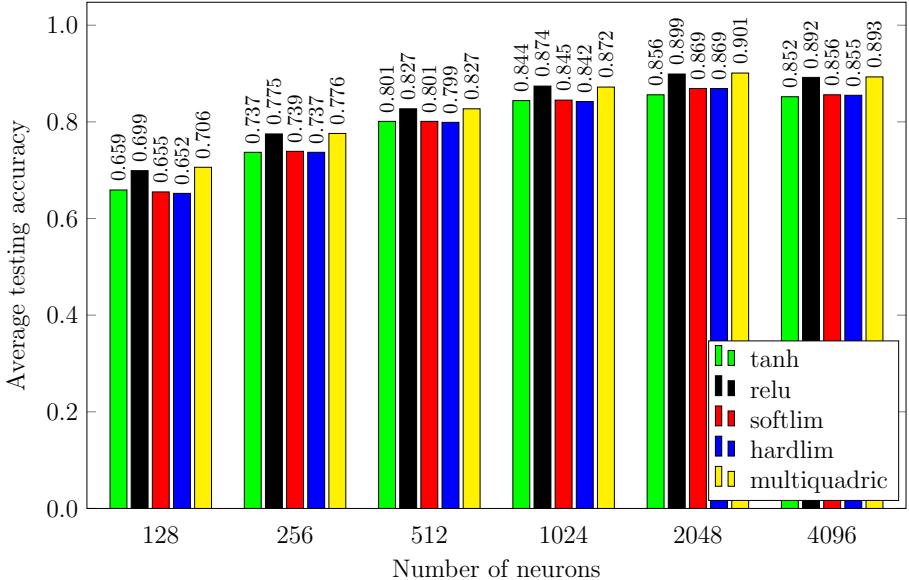

\centering
\input figures/fig_act_elm_1.tex
\caption{Average accuracy of~50 ELM models ($\alpha = 1$)}\label{fig:elm_1_test_bar}
\end{figure}

At the hidden layer, we experiment with
the activation functions \texttt{tanh}, \texttt{relu}, \texttt{softlim}, \texttt{hardlim}, \texttt{multiquadric},
and with number of neurons selected from
$$
  (128, 256, 512, 1024, 2048, 4096) .
$$
Figure~\ref{fig:elm_1_test_bar} shows the average accuracies over~50 ELMs 
with the specified parameters. 

%\begin{figure}[!htb]
%\centering
%\includegraphics[scale=0.4]{images/elm_1_test_bar.png}
%\caption{Average accuracy of 50 ELMs ($\alpha = 1$)}\label{fig:elm_1_test_bar}
%\end{figure}

From the results in Figure~\ref{fig:elm_1_test_bar},
is is clear that the \texttt{relu} and \texttt{multiquadric} activation functions 
consistently perform somewhat better than \texttt{tanh}, 
\texttt{softlim}, and \texttt{hardlim}. Moreover, the average accuracies 
increase with more neurons in the hidden layer, 
until a slight drop from~$2048$ to~$4096$ neurons. 
%%%%% The following is not shown in any graph ?????
%At~$4096$ neurons, the training accuracy reaches~$1.0$ 
%across all activation functions tested.
%This drop in accuracy, along with the trend in training accuracy 
%points to ELMs overfitting when the number
%of neurons reaches~4096

Despite the fact that the weights~$W$ 
and biases~$B$ are selected at random, the ELMs 
in these experiments perform surprisingly consistently.
The standard deviations in the testing accuracies 
are given in Table~\ref{tab:ELM_alpha_1_sd},
and serve to confirm a high degree of consistency,
regardless of the activation function.

\begin{table}[!htb]
\begin{center}
\caption{Standard deviation over~50 ELM models ($\alpha=1$)}\label{tab:ELM_alpha_1_sd}
\begin{tabular}{c|cccccc}\midrule\midrule
\multirow{2}{*}{\bf Activation} & \multicolumn{6}{c}{\bf Neurons}\\
 & 128 & 256 & 512 & 1024 & 2048 & 4096 \\ \hline 
tanh & 0.014 & 0.013 & 0.010 & 0.008 & 0.096 & 0.007\\
relu & 0.013 & 0.011 & 0.008 & 0.007 & 0.007 & 0.008\\
softlim & 0.016 & 0.015 & 0.009 & 0.009 & 0.009 & 0.009\\
hardlim & 0.016 & 0.011 & 0.008 & 0.007 & 0.008 & 0.007\\
multiquadric & 0.012 & 0.011 & 0.007 & 0.008 & 0.005 & 0.006\\
\midrule\midrule 
\end{tabular}
\end{center}
\end{table}

\begin{figure}[!htb]
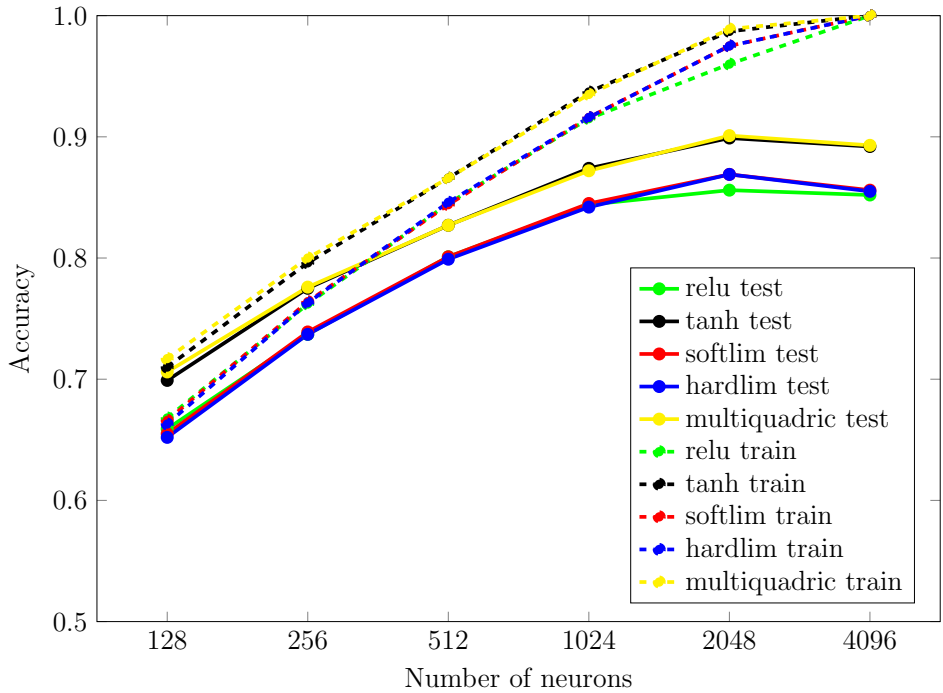

\centering
\input figures/elm_1_train_test.tex
\caption{Training/testing accuracy vs number of neurons ($\alpha=1$)}\label{fig:elm_1_train_test}
\end{figure}

In Figure~\ref{fig:elm_1_train_test} we compare training and testing accuracies
as the number of neurons increases. In each of these experiments, we 
let~$\alpha=1$ and the results are an average of~50 ELMs.
We observe that the training accuracy reaches~1.0 with~4096 neurons,
and this holds for all of the activation functions tested.
However, in every case, the test (or validation) accuracy declines
at~4096 neurons, as compared to~2048 neurons. This is a clear sign
of the models overfitting when~4096 neurons are used.

%\begin{figure}[!htb]
%\centering
%\includegraphics[scale=0.4]{images/elm_1_train_test}
%\caption{Training and testing accuracy as a function of the number of neurons (50 ELMs, $\alpha=1$)} 
%\label{fig:elm_1_train_test}
%\end{figure}

\begin{figure}[!htb]
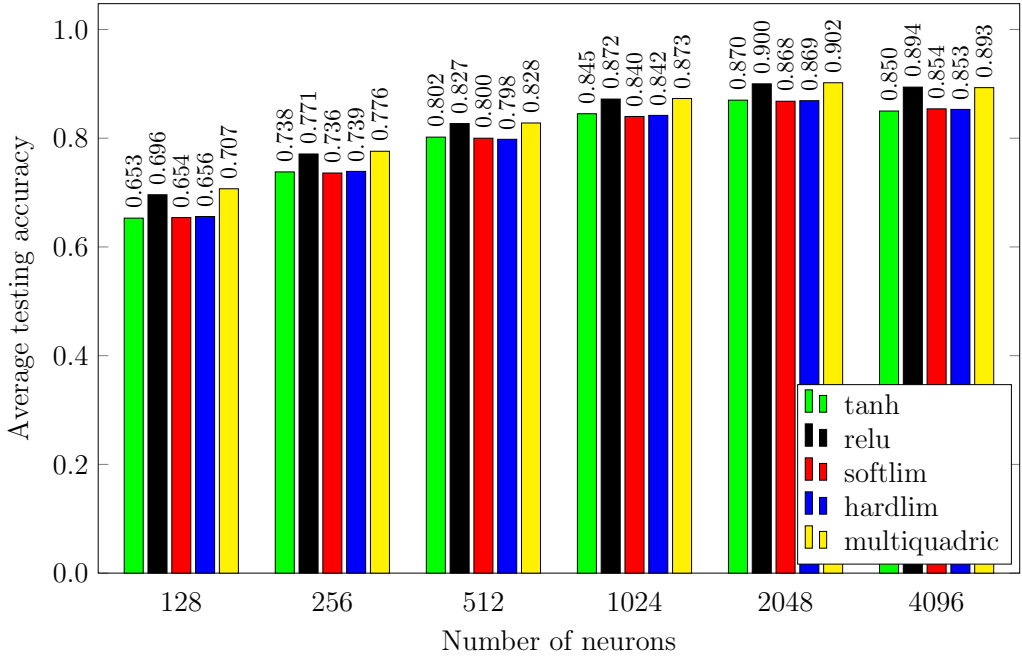

\centering
\input figures/fig_act_elm_05.tex
\caption{Average accuracy of~50 ELM models ($\alpha = 0.5$)}\label{fig:elm_5_test_bar}
\end{figure}

Next, we generate~50 ELMs with the same parameters as above, except
the we use~$\alpha=0.5$ instead of~$\alpha=1$. In this case 
input activation is equally split between the MLP and RBF kernels.
Figure~\ref{fig:elm_5_test_bar} shows the average accuracies across 
these~50 ELM models for the indicated parameters.

%\begin{figure}[!htb]
%\centering
%\includegraphics[scale=0.4]{images/elm_5_test_bar.png}
%\caption{Average accuracy of~50 ELM models ($\alpha = 0.5$)}\label{fig:elm_5_test_bar}
%\end{figure}

We see that the accuracies for the~$\alpha=0.5$ case,
as summarized in Figure~\ref{fig:elm_5_test_bar},
are virtually indistinguishable from the results for the~$\alpha=1$ 
case as given in Figure~\ref{fig:elm_1_test_bar}. In both cases,
the \texttt{relu} and \texttt{multiquadric} activation functions
consistently outperform \texttt{tanh}, \texttt{softlim}, and \texttt{hardlim}. 

\begin{figure}[!htb]
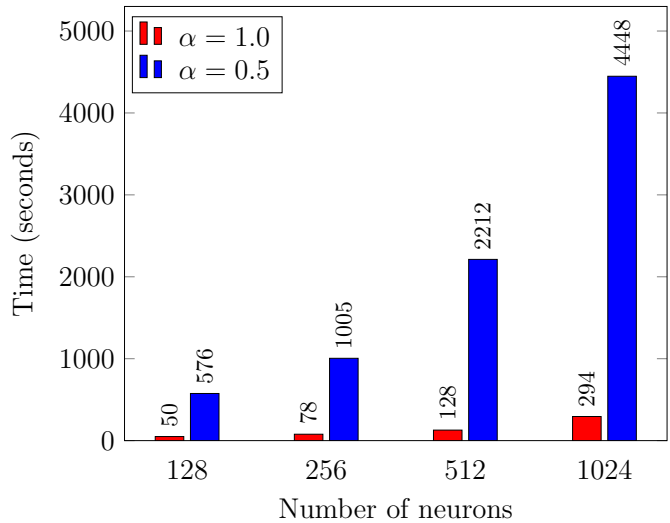

\centering
\input figures/fig_elm_train_time.tex
\caption{Training time for~50 ELMs}\label{fig:elm_train_time}
\end{figure}

We also conducted experiments with a pure RBF kernel, i.e., $\alpha=0$,
but the results were no better than the~$\alpha=0.5$ case. Furthermore, for
any~$0\leq\alpha < 1.0$, the training time is much greater than for the~$\alpha=0$
case. This is clearly illustrated in Figure~\ref{fig:elm_train_time},
where timings for~$\alpha=0$ and~$\alpha=0.5$ are given.

%Next, we consider ELMs trained on~$64\times 64$ images 
%with $\alpha=0$, which represents a pure RBF kernel input activation. 
%As above, we experiment with the activation functions
%\texttt{tanh}, \texttt{relu}, \texttt{softlim}, \texttt{hardlim}, \texttt{multiquadric}.
%%%%%% Why not 2048 and 4096 too ?????
%%%%%% Seems strange to omit cases for super-efficient ELM cases ?????
%Due to the increased training times,
%in these experiments, the number of neurons 
%considered is $(128, 256, 512, 1024)$. 
%Figure~\ref{fig:elm_0_test_bar} shows the average accuracies across~50 ELMs 
%for each case.

%%\begin{figure}[!htb]
%%\centering
%%\includegraphics[scale=0.4]{images/elm_0_test_bar.png}
%%\caption{Average accuracy of 50 ELMs ($\alpha = 0$)}\label{fig:elm_0_test_bar}
%%\end{figure}

%%%%%% results look very suspect with lots of 0.326 ?????
%\begin{figure}[!htb]
%\centering
%\input figures/fig_act_elm_0.tex
%\caption{Average accuracy of 50 ELMs ($\alpha = 0$)}\label{fig:elm_0_test_bar}
%\end{figure}

%As above, \texttt{relu} and \texttt{multiquadric} outperform other activation functions 
%and accuracies increase with the number of neurons. The performance 
%of \texttt{tanh}, \texttt{softlim} and \texttt{hardlim} is poor,
%and shows no improvement with increasing number of neurons.  
%The results for \texttt{relu} and \texttt{multiquadric} are essentialy the same 
%as compared to the~$\alpha=1$ case, above.

The~$\alpha=1$ case is the most efficient and it
performs as well as the~$\alpha=0.5$ case. We also find that~1024
neurons performs best, and that the~\texttt{relu}
activation outperforms the other activations. Hence, we
have selected the parameters in Table~\ref{tab:ELM_parms}
for all ELM experiments discussed below, unless explicitly stated otherwise.

\begin{table}[!htb]
\begin{center}
\caption{ELM parameters}\label{tab:ELM_parms}
\begin{tabular}{c|c|c}\midrule\midrule
%\multirow{2}{*}{\bf Activation} & \multicolumn{6}{c}{\bf Neurons}\\
 \textbf{Activation} & \textbf{Mixing} & \textbf{Neurons} \\ \hline
\texttt{relu} & $\alpha=1$ & $1024$\\
\midrule\midrule 
\end{tabular}
\end{center}
\end{table}
%$$
%    \mbox{Activation: \texttt{relu}},\  
%    \mbox{Mixing: $\alpha=1$},\ 
%    \mbox{Neurons: $1024$}
%$$ 
%\begin{itemize}
%    \item Activation function: \texttt{relu} 
%    \item Mixing parameter: $\alpha=1$
%    \item Number of neurons: $1024$
%\end{itemize} 

\subsubsection{Majority Vote Ensembles}

Ensemble learning consists of combining several models into one,
which can serve to decrease the variance and the bias, thus improving predictions. 
In this section, we consider a ensemble technique consisting
of a majority vote of classifiers. Here, the individual classifiers are~50 ELMs 
and, again, these are combined by taking a simple majority vote of their decisions. 
Since we have~25 classes, it is possible that there could be ties in
the voting, in which case we would simply make a random selection
from among those receiving the most votes. However, we found
no case where there was a tie vote in any of our experiments.

\begin{figure}[!htb]
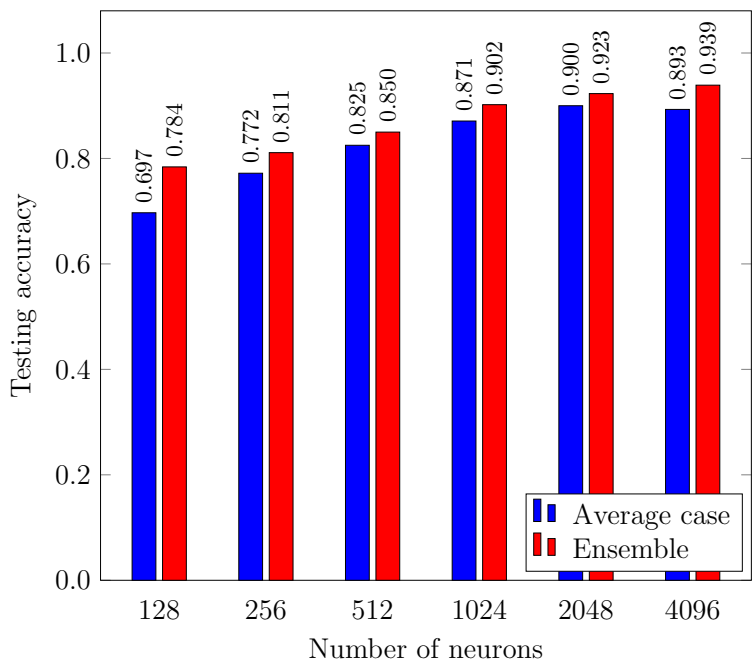

\centering
\input figures/fig_elm_ensemble.tex
\caption{Ensemble classifier accuracy ($64\times 64$ images)}\label{fig:elm_ensemble}
\end{figure}

Figure~\ref{fig:elm_ensemble} shows that with increasing number of neurons, 
our ensemble accuracy consistently increases. Furthermore, the ensemble classifier 
always outperforms the average case for the individual ELMs.%%%%% best case ?????

%\begin{figure}[!htb]
%\centering
%\includegraphics[scale=0.5]{images/elm_ensemble.png}
%\caption{Ensemble classifier accuracy ($64\times 64$ images)}\label{fig:elm_ensemble}
%\end{figure}

\subsubsection{Dropouts}

\begin{figure}[!htb]
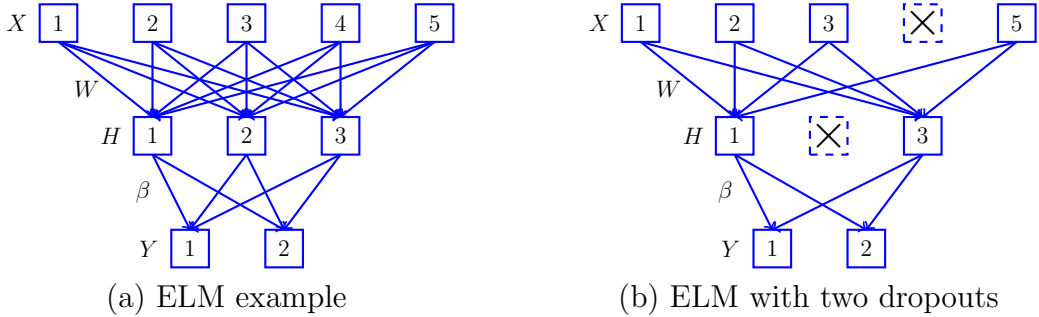

  \centering
  \begin{tabular}{ccc}
    \input figures/elm2.tex
    &
    \hspace*{0.33in}
    &
    \input figures/elm2_drop.tex
    \\
    (a) ELM example
    &
    &
    (b) ELM with two dropouts
    \end{tabular}
  \caption{Dropouts}\label{fig:dropout}
\end{figure}

In a fully connected layer, each neuron is connected to all input nodes. 
Dropouts are neurons that are ignored during the training phase---an example
appears in Figure~\ref{fig:dropout}. The neurons to be dropped are chosen at random with a 
probability~$p$, with typically only a small percentage of neurons 
are dropped~\cite{dropoutimage}.
Dropouts offer a simple form of regularization, and hence serve to reduce overfitting.

Since ELMs have no backpropagation, we take a slightly different approach 
to dropouts in our ELM experiments. We implement an algorithm that 
connects each neuron in the hidden layer to four randomly selected
input nodes. This technique results
in a relatively large percentage of connections being lost.
Such an aggressive approach is needed with ELMs, since training
occurs in a single step, as opposed to multiple epochs
in backpropagation.

\begin{figure}[!htb]
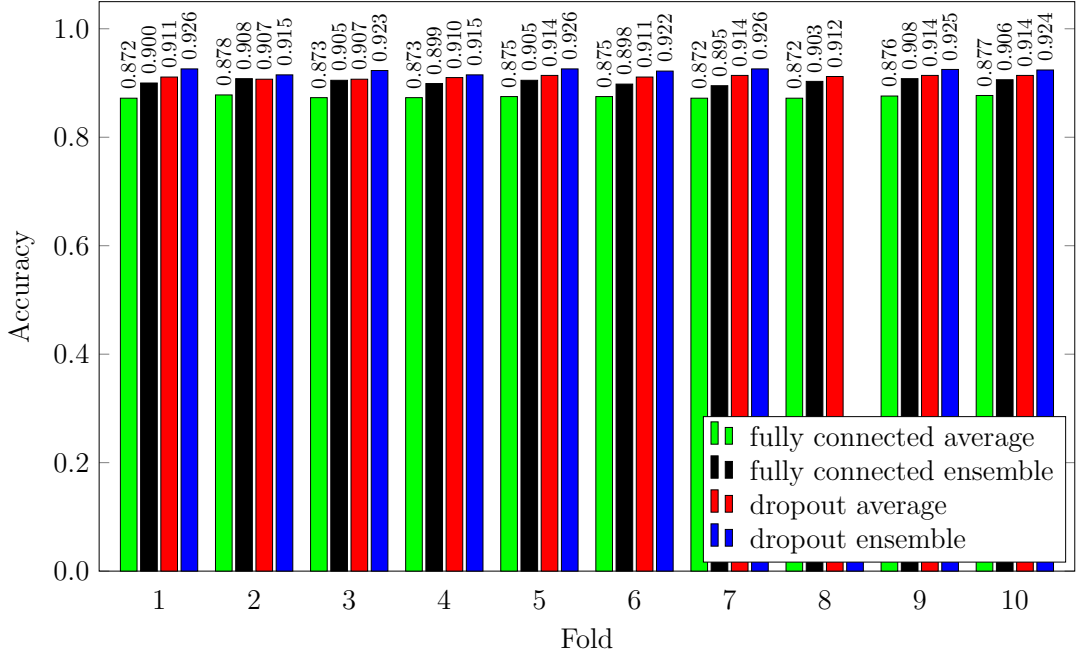

\centering
\input figures/fig_elm_dropout.tex
\caption{Fully connected vs dropout ELMs ($64\times 64$ images)}\label{fig:full_dropout_64}
\end{figure}

Figure~\ref{fig:full_dropout_64} gives our average and ensemble accuracies 
across~50 ELMs, for both the fully connected and dropout cases. 
Dropout ELMs consistently perform better than fully connected ELMs. 
In addition, the corresponding ensemble classifiers
consistently outperform the corresponding non-ensemble classifiers.

%\begin{figure}[!htb]
%\centering
%\includegraphics[scale=0.4]{images/full_dropout_64_avg.png}
%\caption{Average accuracy for fully connected vs dropout ELMs ($64\times 64$ images)} 
%\label{fig:full_dropout_64_avg}
%\end{figure}

%\begin{figure}[!htb]
%\centering
%\includegraphics[scale=0.4]{images/full_dropout_64_ens.png}
%\caption{Ensemble accuracy for fully connected vs dropout ELMs ($64\times 64$ images)} 
%\label{fig:full_dropout_64_ens}
%\end{figure}

\subsubsection{One Dimensional Input}

An executable file can be viewed as a sequence of bytes. When 
executable files are converted to images, sequential patterns present 
in the executables may be lost, depending on the image dimensions. 
To preserve these sequential patterns, we consider a different technique for 
generating ELM input feature vectors. Instead of treating the Malimg 
samples as images, we consider each as a one-dimensional vector
by simply reading the pixels from left to right from top to bottom. 
Since the images in the dataset are of varying dimensions, 
these one-dimensional vector also differs in length.

We convert all of out one-dimensional vectors to a fixed length~$N$ 
by simply averaging elements. Specifically, 
suppose that a vector~$V$ is of length~$T$. If necessary, we pad~$V$
so that it is evenly divisible by~$N$. Let~$V'$ be this new vector, and
let~$T'$ be its length. Then we generate a vector~$X$ of length~$N$ where
$$
  x_i = \frac{v_i' + v_{i+1}' + \cdots + v_{i+N-1}'}{\ell}
$$
where~$\ell = T'/N$.
We refer the resulting~$X$ as an~$N\times 1$ vectors
to emphasize that it is 1-dimensional, as opposed to the
2-dimensional image from which it was derived.
These~$N\times 1$ vectors serve as the features for all 
experiments in this section.

\begin{figure}[!htb]
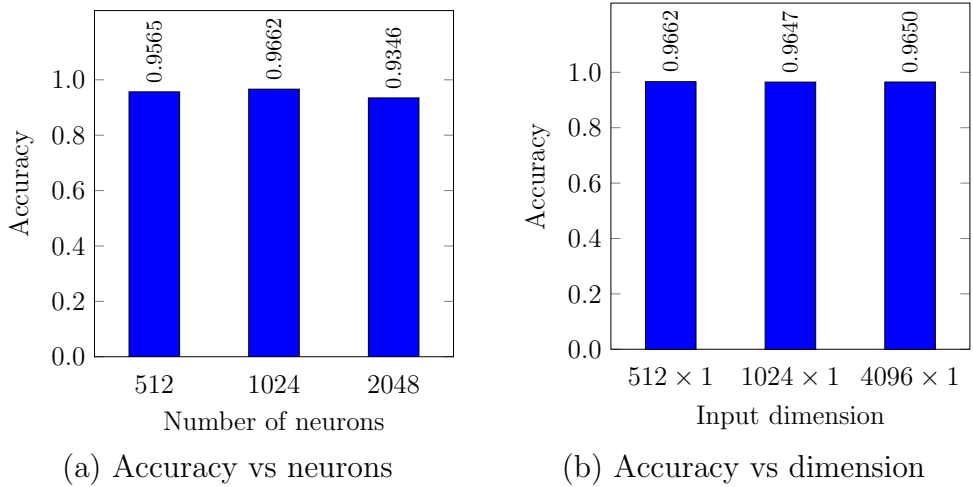

\centering
\begin{tabular}{ccc}
\input figures/fig_elm_a.tex
& &
\input figures/fig_elm_b.tex
\\
(a) Accuracy vs neurons
& &
(b) Accuracy vs dimension
\end{tabular}
\caption{Average ensemble accuracy}\label{fig:elm_ab}
\end{figure}

Our first one-dimensional input experiment 
consists of ensemble classifiers comprised of~$50$ dropout ELMs. 
The input to each ELM is a vector of dimension~$512\times 1$. 
We train one such ensemble ELM with~$512$ neurons, another with~$1024$, 
and a third with~$2048$ neurons. In each case, we use~$\alpha = 1$
and \texttt{relu} activation functions.  
Figure~\ref{fig:elm_ab}~(a) shows the average ensemble classifier accuracy,
and we see that the best accuracy is achieved for ELMs with~1024 neurons.

%\begin{figure}[!htb]
%\centering
%\includegraphics[scale=0.5]{images/elm_a.png}
%\caption{Average ensemble accuracy vs number of neurons ($512\times 1$ input}\label{fig:elm_a}
%\end{figure}

We consider an additional set of experiments using~1024 neurons and the same 
parameters as in our previous experiment, but varying the input vector length.
Figure~\ref{fig:elm_ab}~(b) shows increasing the input vector length beyond~$512\times 1$
has a minimal effect on the accuracy.
Interestingly, we see little variability in these one-dimensional input length results.
This is in contrast to the CNN results on 2-dimensional images, where 
the image size was a significant parameter.

%\begin{figure}[!htb]
%\centering
%\includegraphics[scale=0.5]{images/elm_b.png}
%\caption{Average ensemble accuracy with varying input dimension}\label{fig:elm_b}
%\end{figure}

%\subsubsubsection{Cross Validation of ELM with Dropout}

%This experiment shows the stability of ELMs specifically with respect to 
%a $1$~dimensional input. 50 dropout ELMs were trained on ten folds of the data. 
%The ELM parameters are  $\alpha=1$ , $1024$ neurons, \texttt{relu} activation function. 

%The input dimension is $512\times 1$. Figure~\ref{fig:elm_c_512} shows the 
%average accuracy of the ensemble classifier over ten folds of the data. The 
%accuracies are consistent across each fold, with the minimum and maximum 
%accuracies as $95.6\%$ and $97.4\%$ respectively. The consistency in 
%accuracies also reinforces the stability of ELMs with respect to the $1$-dimensional input method.

%\begin{figure}[!htb]
%\centering
%\includegraphics[scale=0.4]{images/elm_c_512.png}
%\caption{Cross validation of dropout ELM (input size = $512\times 1$)}\label{fig:elm_c_512}
%\end{figure}

\subsubsection{Weighted ELMs}\label{sect:weightedELM}

The Malimg dataset is highly imbalanced, as can be seen from
the numbers in Table~\ref{tab:families}. In a classification task with such unequal 
number of samples per class, predictions are naturally biased towards 
the classes with the most data. To better account for imbalanced 
class distributions, a weighted ELM is proposed in~\cite{weightedelm},
which employs a weighted linear system for the solution of the output layer weights. 
To compute the necessary weights, we let
$$
  S_j = \text{total number of samples in class } j
$$
and
$$
  S = \sum S_j
$$
where the sum is over all classes in the dataset.
Denote row~$j$ of~$H$ as~$h_j$
and row~$j$ of~$Y$ as~$y_j$
and let~$c_j=\sqrt{S/S_j}$. Then we compute
$$
  h'_j = c_j h_j \mbox{\ \ and\ \ } y'_j = c_j y_j
$$
and form the weighted matrices~$H'$ and~$Y'$
from these weighted rows. The weighted ELM is trained
as discussed in Section~\ref{sect:ELMs}, except that
we use~$H'$ and~$Y'$ and place of~$H$ and~$Y$.

\begin{figure}[!htb]
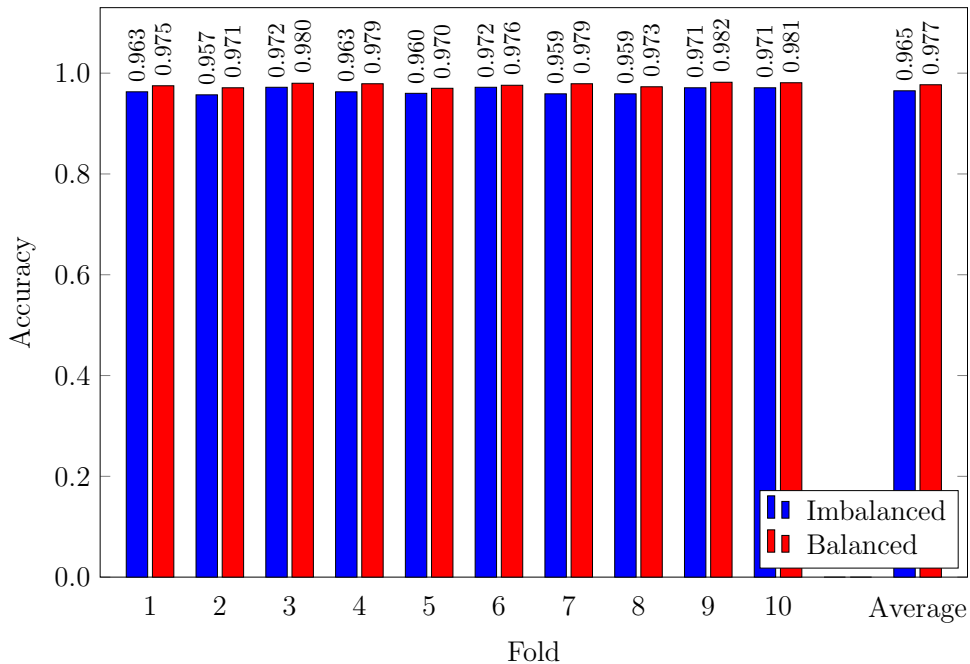

\centering
\input figures/fig_elm_weighted.tex
\caption{Weighted ELM accuracy ($1024\times 1$ input)}\label{fig:elm_d}
\end{figure}

For our weighted ELM experiments, we consider an ensemble of~50 dropout ELMs. 
The individual ELMs have~$1024$ neurons, and we use~$\alpha=1$
with \texttt{relu} activation functions, and one-dimensional input
vectors of size~$1024\times 1$. Figure~\ref{fig:elm_d} compares the accuracy 
of the weighted and unweighted ensemble classifiers. The average accuracy for the 
unweighted model is~96.5\%, while for the weighted model we obtain
a slight improvement at~97.7\%. 

%\begin{figure}[!htb]
%\centering
%\includegraphics[scale=0.4]{images/elm_d.png}
%\caption{Weighted ELM accuracy ($1024\times 1$ input)}\label{fig:elm_d}
%\end{figure}

\section{Discussion}

\begin{figure}[!htb]
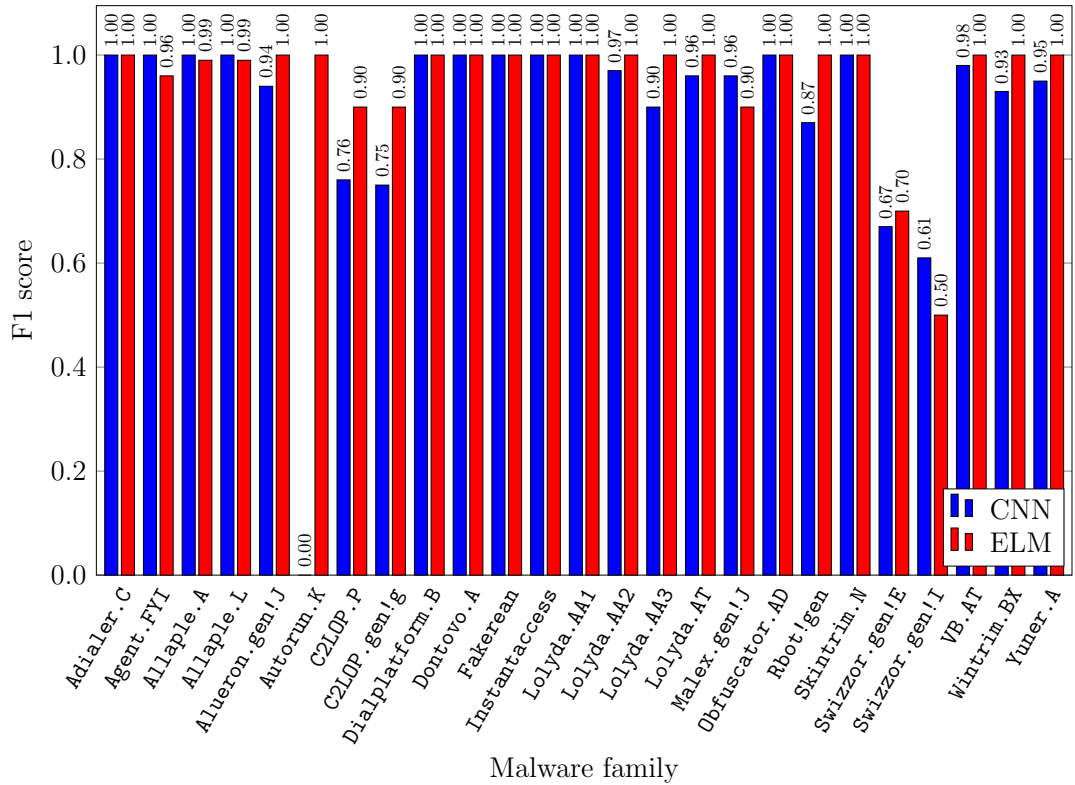

\centering
\input figures/fig_f1.tex
\caption{Classwise F1 scores for CNN and ELM}\label{fig:classwise_f1_comparison}
\end{figure}

Our best CNN model achieves an overall accuracy of~96.3\%. 
This CNN is based on an input image of size~$128\times 128$, it uses
one convolutional layer with~$32$ filters, each of size~$3\times 3$,
a max-pooling layer with a filter of size~$2\times 2$, 
a dense layer of~$128$ neurons, and~$25$ neurons in the output layer.
In comparison, our best ELM-based model achieves an accuracy of~97.7\%. 
This ELM-based model consists of an ensemble classifier built 
on~50 ELMs with dropouts, each of which
uses a class weighting technique, with~$\alpha = 1$ 
and~$1024$ neurons in the hidden layer. These ensembled ELMs
all use \texttt{relu} activation functions and are trained on 
one-dimensional input of size~$1024\times 1$. 
Figure~\ref{fig:classwise_f1_comparison} gives
the classwise~F1 score comparison for our best CNN and ELM models.

%\begin{figure}[!htb] 
%\centering
%\includegraphics[scale=0.4]{images/classwise_f1_comparison}
%\caption{Classwise F1 scores for CNN and ELM}\label{fig:classwise_f1_comparison}
%\end{figure}

In addition to outperforming CNNs overall, our
ELM achieves equal or better~F1 scores on~18 of the~25 classes. 
It is interesting to note that the ELMs are also able to 
predict at least some instances of all~25 families, 
whereas the CNN sometimes misses an entire family. 
For example, in Figure~\ref{fig:classwise_f1_comparison}, we see
that the CNN misclassifies all instances of the \texttt{Autorun.K} family,
while in stark contrast, the ELM is able to correctly classify all 
samples from this particular family. 

\begin{figure}[!htb]
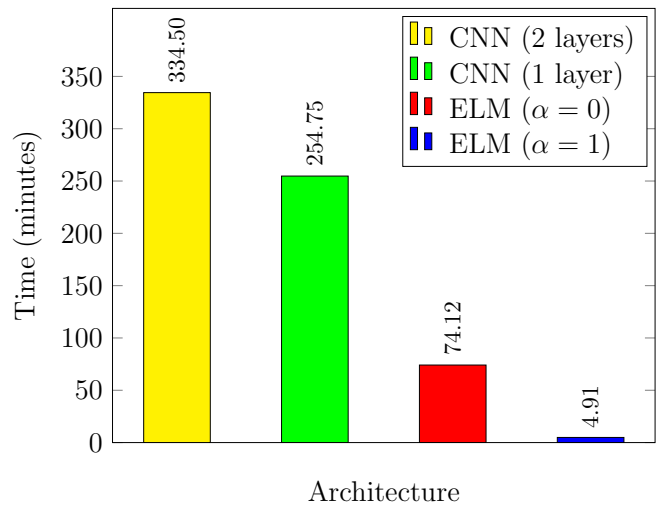

\centering
\input figures/fig_cnn_elm_time2.tex
\caption{Training time comparison of CNN and ELM}\label{fig:cnn_elm_time}
\end{figure}

Finally, we consider the training efficiency of ELMs in comparison to CNNs. 
In Figure~\ref{fig:cnn_elm_time} we give timings for some of the architectures
discussed above. In each case, the time listed is for training~50 models of the
specified type. 

We see that standard ELMs with the MLP kernel (i.e., the~$\alpha = 1.0$ case)
can be trained in about~2\%\ of the time required to train a CNN having just a
single convolution layer. In comparison to 2-convolutional layers, we
can train a standard ELM in less than~1.5\%\ of the time required for the CNN.
And, even with the computationally expensive RBF kernel, the ELM can be trained in
less than~30\%\ of the time required by the fastest CNN architecture. Given
that we obtained our best results with an ELM using the MLP kernel, these
results show a clear advantage of ELMs in cases where a large number of models
are to be trained.

%\section{Conclusion and Future Work}\label{chap:conclusion}
\section{Conclusion}\label{chap:conclusion}

In this paper, we have compared CNNs and ELMs for image-based malware classification. 
We considered a variety of different CNN architectures, varying the input image size, 
number of convolutional layers, and number of filters. In another set of experiments 
we tested the performance of a wide variety of ELM architectures and parameters, 
using both two-dimensional images one-dimensional vectors derived from these images.

Although CNNs are widely used for image classification, we found that,
based on a substantial number of experiments, 
ELMs were able to outperform CNNs on the well-known Malimg 
dataset. The primary advantage of ELMs is faster training times
as compared to CNNs. %%%%% results????? (as shown in Appendix~\ref{app:a}?????). 
In our experiments, one CNN model typically required hours to train, while
one ELM could be trained in a few seconds. This makes ELMs 
an option that should be considered for malware classification and
detection, due to the large volume of new malware and variants released each year. 
ELMs have many detractors, but our experimental results strongly indicate
that they can be competitive in the malware domain, at least in some situations.

For future work, it would be interesting to consider malware classification
within broad types (trojan, worm, backdoor, etc.), rather than specific families.
Experiments could also be conducted on other malware datasets 
and based on other techniques for converting executables to images. Our one-dimensional
feature vector experiments suggest that for ELMs, it might not be necessary to consider
images at all, and hence additional experiments with raw executable files would be worthwhile.
Finally, many additional combinations of parameters could be considered,
both for CNNs and ELMs.

\bibliographystyle{plain}
\bibliography{references.bib}

%
% ***** appendices in appA.tex, appB.tex, ...
%
%\newappendix{appA}

\end{document}

%% file: preamble.tex
\usepackage{amsmath,amsthm, amsfonts, amssymb, amsxtra,amsopn}
\usepackage{pgfplots}
\usepgfplotslibrary{colormaps}
\pgfplotsset{compat=1.15}
\usepackage{pgfplotstable}
\usetikzlibrary{pgfplots.statistics} 
\usepackage{filecontents}
\usepackage{graphicx}
\usepackage{multirow}
\usepackage{tabu}
\usepackage{algorithmic}
\usepackage{longtable}
\usepackage{booktabs}
\usepackage{listings}
\usepackage{cmap}
\usepackage{colortbl}
\usepackage{adjustbox}
\usepackage{epsfig}
\usepackage{xstring}%%% needed for \rowvec macros

\usepackage{subfigure}

\PassOptionsToPackage{hyphens}{url}

\usepackage{hyperref}
\hypersetup{colorlinks=true,linkcolor=black,citecolor=black,urlcolor=blue,filecolor=black}
\hypersetup{pdfpagemode=UseNone,pdfstartview=}

\usepackage[tableposition=top]{caption}
\usepackage{enumitem}
\setlist[itemize]{noitemsep, topsep=0pt}

\advance\oddsidemargin by -0.45in
\advance\textwidth by 0.9in

\advance\topmargin by -0.3in
\advance\textheight by 0.6in

\def\eref#1{(\ref{#1})}

\long\def\symbolfootnotetext[#1]#2{\begingroup%
\def\thefootnote{\fnsymbol{footnote}}\footnotetext[#1]{#2}\endgroup}

\DeclareMathOperator{\thth}{th}

\newcommand{\bftab}{\fontseries{b}\selectfont}

\def\z{\phantom{0}}

\def\t{\footnotesize}
\newcommand{\mThree}{$\scalebox{0.7}[1.0]{$-$} 3$}
\newcommand{\mTwo}{$\scalebox{0.7}[1.0]{$-$} 2$}
\newcommand{\mOne}{$\scalebox{0.7}[1.0]{$-$} 1$}

\def\rbfKernel{R}
\def\mlpKernel{M}

\pgfkeys{
    /pgf/number format/fixed zerofill=true }
    
\pgfplotstableset{
    color cells/.code={%
        \pgfqkeys{/color cells}{#1}%
        \pgfkeysalso{%
            postproc cell content/.code={%
                \begingroup
                \pgfkeysgetvalue{/pgfplots/table/@preprocessed cell content}\value
\ifx\value\empty
\endgroup
\else
                \pgfmathfloatparsenumber{\value}%
                \pgfmathfloattofixed{\pgfmathresult}%
                \let\value=\pgfmathresult
                \pgfplotscolormapaccess
                    [\pgfkeysvalueof{/color cells/min}:\pgfkeysvalueof{/color cells/max}]%
                    {\value}%
                    {\pgfkeysvalueof{/pgfplots/colormap name}}%
                \pgfkeysgetvalue{/pgfplots/table/@cell content}\typesetvalue
                \pgfkeysgetvalue{/color cells/textcolor}\textcolorvalue
                \toks0=\expandafter{\typesetvalue}%
                \xdef\temp{%
                    \noexpand\pgfkeysalso{%
                        @cell content={%
                            \noexpand\cellcolor[rgb]{\pgfmathresult}%
                            \noexpand\definecolor{mapped color}{rgb}{\pgfmathresult}%
                            \ifx\textcolorvalue\empty
                            \else
                                \noexpand\color{\textcolorvalue}%
                            \fi
                            \the\toks0 %
                        }%
                    }%
                }%
                \endgroup
                \temp
\fi
            }%
        }%
    }
}

\def\srowvecc#1#2{(\!\begin{array}{cc} 
      \noexpandarg\IfBeginWith{#1}{-}{\! #1}{#1}
    & #2\kern-0.5pt\end{array}\!)}
\def\rowvecc#1#2{\left(\!\begin{array}{cc} 
      \noexpandarg\IfBeginWith{#1}{-}{\! #1}{#1}
    & #2\kern-0.5pt\end{array}\!\right)}
\def\rowveccc#1#2#3{\left(\!\begin{array}{ccc} 
      \noexpandarg\IfBeginWith{#1}{-}{\! #1}{#1}
    & #2 
    & #3\kern-0.5pt\end{array}\!\right)}
\def\rowvecccc#1#2#3#4{\left(\!\begin{array}{cccc}
      \noexpandarg\IfBeginWith{#1}{-}{\! #1}{#1}
    & #2 
    & #3 
    & #4\kern-0.5pt\end{array}\!\right)}
\def\srowvecccc#1#2#3#4{\bigl(\!\begin{array}{cccc}
      \noexpandarg\IfBeginWith{#1}{-}{\! #1}{#1}
    & #2 
    & #3 
    & #4\kern-0.5pt\end{array}\!\bigr)}
\def\rowveccccc#1#2#3#4#5{\left(\!\begin{array}{ccccc} 
      \noexpandarg\IfBeginWith{#1}{-}{\! #1}{#1}
    & #2
    & #3
    & #4
    & #5\kern-0.5pt\end{array}\!\right)}
\def\srowvecccccc#1#2#3#4#5#6{(\!\begin{array}{cccccc} 
      \noexpandarg\IfBeginWith{#1}{-}{\! #1}{#1}
    & #2
    & #3
    & #4
    & #5
    & #6\kern-0.5pt\end{array}\!)}
\def\rowvecccccc#1#2#3#4#5#6{\left(\!\begin{array}{cccccc} 
      \noexpandarg\IfBeginWith{#1}{-}{\! #1}{#1}
    & #2
    & #3
    & #4
    & #5
    & #6\kern-0.5pt\end{array}\!\right)}

%% file: figures/conv3b.tex
%    \begin{tikzpicture}[thick,scale=0.9]
\begin{tikzpicture}[scale=0.4]

% grid 1
%\draw[step=0.7,red,very thin] (0.0,0.0) grid (11.2,11.2);
\draw[red,ultra thick] (0.0,0.0) rectangle (11.2,11.2);

% grid 2
\draw[green,ultra thick] (0.5,-0.5) rectangle (11.7,10.7);
%\draw[step=0.7,green,very thin, shift={(0.5,-0.5)}] (0.0,0.0) grid (11.2,11.2);

% grid 3
\draw[blue,ultra thick] (1.0,-1.0) rectangle (12.2,10.2);
%\draw[step=0.7,blue,very thin, shift={(1.0,-1.0)}] (0.0,0.0) grid (11.2,11.2);

%% grid
%\draw[step=0.7,black,thin] (12.59,1.39) grid (22.4,11.2);
%\draw[black,ultra thick] (12.6,1.4) rectangle (22.4,11.2);

%% convolution 1
\draw[ultra thick, dotted] (0.0,9.1) rectangle (2.1,11.2);
\draw[ultra thick] (0.0,11.2) -- (2.1,11.2);
\draw[ultra thick] (0.0,11.2) -- (0.0,9.1);
\draw[ultra thick] (1.0,8.1) rectangle (3.1,10.2);
\draw[ultra thick] (0.0,9.1) -- (1.0,8.1);
\draw[ultra thick] (2.1,11.2) -- (3.1,10.2);
\draw[ultra thick] (0.0,11.2) -- (1.0,10.2);
\draw[ultra thick, dotted] (3.1,8.1) -- (2.1,9.1);

\draw[black,ultra thick] (13.6,1.4) rectangle (23.4,11.2);
\draw[ultra thick] (13.6,11.2) rectangle (14.3,10.5);
%\draw[step=0.7,gray,very thin, shift={(13.6,1.4)}] (0.0,0.0) grid (9.8,9.8);
\draw[black,ultra thick] (14.1,0.9) rectangle (23.9,10.7);
\draw[ultra thick] (14.1,10.7) rectangle (14.8,10.0);
%\draw[step=0.7,gray,very thin, shift={(14.1,0.9)}] (0.0,0.0) grid (9.8,9.8);
\draw[black,ultra thick] (14.6,0.4) rectangle (24.4,10.2);
\draw[ultra thick] (14.6,10.2) rectangle (15.3,9.5);
%\draw[step=0.7,gray,very thin, shift={(14.6,0.4)}] (0.0,0.0) grid (9.8,9.8);
\draw[black,ultra thick] (15.1,-0.1) rectangle (24.9,9.7);
\draw[ultra thick] (15.1,9.7) rectangle (15.8,9.0);
%\draw[step=0.7,gray,very thin, shift={(15.1,-0.1)}] (0.0,0.0) grid (9.8,9.8);
\draw[black,ultra thick] (15.6,-0.6) rectangle (25.4,9.2);
\draw[ultra thick] (15.6,9.2) rectangle (16.3,8.5);
%\draw[step=0.7,gray,very thin, shift={(15.6,-0.6)}] (0.0,0.0) grid (9.8,9.8);

%% line
\draw[smooth,thick,->] (2.5,10.7) -- (2.5,11.9) -- (3.0,12.2) -- (12.95,12.2)  -- (13.55,11.9) -- (13.95,11.2);
\node at (7,12.7) {$\mbox{\footnotesize filter}_1$};
%% line
\draw[smooth,thick,->] (2.0,10.7) -- (2.0,12.9) -- (2.5,13.2) -- (12.95,13.2)  -- (13.55,12.9) -- (14.45,10.7);
\node at (7,13.7) {$\mbox{\footnotesize filter}_2$};
%% line
\draw[smooth,thick,->] (1.5,10.7) -- (1.5,13.9) -- (2.0,14.2) -- (12.95,14.2)  -- (13.55,13.9) -- (14.95,10.2);
\node at (7,14.7) {$\mbox{\footnotesize filter}_3$};
%% line
\draw[smooth,thick,->] (1.0,10.7) -- (1.0,14.9) -- (1.5,15.2) -- (12.95,15.2)  -- (13.55,14.9) -- (15.45,9.7);
\node at (7,15.7) {$\mbox{\footnotesize filter}_4$};
%% line
\draw[smooth,thick,->] (0.5,10.7) -- (0.5,15.9) -- (1.0,16.2) -- (12.95,16.2)  -- (13.55,15.9) -- (15.95,9.2);
\node at (7,16.7) {$\mbox{\footnotesize filter}_5$};

\end{tikzpicture}

%% file: figures/maxPool.tex
\begin{tikzpicture}[scale=0.675,every node/.style={scale=0.9}]

% grid
\draw[step=0.7,black,thin] (12.59,1.39) grid (22.4,11.2);
\draw[black,ultra thick] (12.6,1.4) rectangle (22.4,11.2);

% row 1
\node at (12.95,10.85) {\t 0};
\node at (13.65,10.85) {\t 0};
\node at (14.35,10.85) {\t 0};
\node at (15.05,10.85) {\t 0};
\node at (15.75,10.85) {\t 0};
\node at (16.45,10.85) {\t 0};
\node at (17.15,10.85) {\t 0};
\node at (17.85,10.85) {\t 0};
\node at (18.55,10.85) {\t 0};
\node at (19.25,10.85) {\t 0};
\node at (19.95,10.85) {\t 0};
\node at (20.65,10.85) {\t 0};
\node at (21.35,10.85) {\t 0};
\node at (22.05,10.85) {\t 0};
% row 2
\node at (12.95,10.15) {\t 0};
\node at (13.65,10.15) {\t 0};
\node at (14.35,10.15) {\t 0};
\node at (15.05,10.15) {\t 2};
\node at (15.75,10.15) {\t 1};
\node at (16.45,10.15) {\t 0};
\node at (17.15,10.15) {\t 0};
\node at (17.85,10.15) {\t 0};
\node at (18.55,10.15) {\t 0};
\node at (19.25,10.15) {\t \mTwo};
\node at (19.95,10.15) {\t \mOne};
\node at (20.65,10.15) {\t 0};
\node at (21.35,10.15) {\t 0};
\node at (22.05,10.15) {\t 0};
% row 3
\node at (12.95,9.45) {\t 0};
\node at (13.65,9.45) {\t 0};
\node at (14.35,9.45) {\t 2};
\node at (15.05,9.45) {\t \mTwo};
\node at (15.75,9.45) {\t 0};
\node at (16.45,9.45) {\t 0};
\node at (17.15,9.45) {\t 0};
\node at (17.85,9.45) {\t 0};
\node at (18.55,9.45) {\t 0};
\node at (19.25,9.45) {\t 3};
\node at (19.95,9.45) {\t \mTwo};
\node at (20.65,9.45) {\t \mOne};
\node at (21.35,9.45) {\t 0};
\node at (22.05,9.45) {\t 0};
% row 4
\node at (12.95,8.75) {\t 0};
\node at (13.65,8.75) {\t 2};
\node at (14.35,8.75) {\t \mTwo};
\node at (15.05,8.75) {\t 0};
\node at (15.75,8.75) {\t \mThree};
\node at (16.45,8.75) {\t 0};
\node at (17.15,8.75) {\t 0};
\node at (17.85,8.75) {\t 0};
\node at (18.55,8.75) {\t 0};
\node at (19.25,8.75) {\t 0};
\node at (19.95,8.75) {\t 6};
\node at (20.65,8.75) {\t \mTwo};
\node at (21.35,8.75) {\t \mOne};
\node at (22.05,8.75) {\t 0};
% row 5
\node at (12.95,8.05) {\t 0};
\node at (13.65,8.05) {\t 1};
\node at (14.35,8.05) {\t 0};
\node at (15.05,8.05) {\t \mThree};
\node at (15.75,8.05) {\t 4};
\node at (16.45,8.05) {\t \mOne};
\node at (17.15,8.05) {\t \mOne};
\node at (17.85,8.05) {\t 2};
\node at (18.55,8.05) {\t \mOne};
\node at (19.25,8.05) {\t \mTwo};
\node at (19.95,8.05) {\t 0};
\node at (20.65,8.05) {\t 3};
\node at (21.35,8.05) {\t \mTwo};
\node at (22.05,8.05) {\t 0};
% row 6
\node at (12.95,7.35) {\t 0};
\node at (13.65,7.35) {\t 0};
\node at (14.35,7.35) {\t 0};
\node at (15.05,7.35) {\t 0};
\node at (15.75,7.35) {\t \mOne};
\node at (16.45,7.35) {\t 2};
\node at (17.15,7.35) {\t \mOne};
\node at (17.85,7.35) {\t \mOne};
\node at (18.55,7.35) {\t 2};
\node at (19.25,7.35) {\t \mOne};
\node at (19.95,7.35) {\t 0};
\node at (20.65,7.35) {\t 0};
\node at (21.35,7.35) {\t 0};
\node at (22.05,7.35) {\t 0};
% row 7
\node at (12.95,6.65) {\t 0};
\node at (13.65,6.65) {\t 0};
\node at (14.35,6.65) {\t 0};
\node at (15.05,6.65) {\t 2};
\node at (15.75,6.65) {\t \mTwo};
\node at (16.45,6.65) {\t \mTwo};
\node at (17.15,6.65) {\t 2};
\node at (17.85,6.65) {\t \mOne};
\node at (18.55,6.65) {\t 1};
\node at (19.25,6.65) {\t 1};
\node at (19.95,6.65) {\t \mOne};
\node at (20.65,6.65) {\t 0};
\node at (21.35,6.65) {\t 0};
\node at (22.05,6.65) {\t 0};
% row 8
\node at (12.95,5.95) {\t 0};
\node at (13.65,5.95) {\t 0};
\node at (14.35,5.95) {\t 0};
\node at (15.05,5.95) {\t \mOne};
\node at (15.75,5.95) {\t 4};
\node at (16.45,5.95) {\t \mTwo};
\node at (17.15,5.95) {\t \mOne};
\node at (17.85,5.95) {\t 2};
\node at (18.55,5.95) {\t \mTwo};
\node at (19.25,5.95) {\t 1};
\node at (19.95,5.95) {\t \mOne};
\node at (20.65,5.95) {\t 0};
\node at (21.35,5.95) {\t 0};
\node at (22.05,5.95) {\t 0};
% row 9
\node at (12.95,5.25) {\t 0};
\node at (13.65,5.25) {\t 0};
\node at (14.35,5.25) {\t 0};
\node at (15.05,5.25) {\t \mOne};
\node at (15.75,5.25) {\t \mTwo};
\node at (16.45,5.25) {\t 6};
\node at (17.15,5.25) {\t 0};
\node at (17.85,5.25) {\t \mThree};
\node at (18.55,5.25) {\t 0};
\node at (19.25,5.25) {\t \mTwo};
\node at (19.95,5.25) {\t 2};
\node at (20.65,5.25) {\t 0};
\node at (21.35,5.25) {\t 0};
\node at (22.05,5.25) {\t 0};
% row 10
\node at (12.95,4.55) {\t 0};
\node at (13.65,4.55) {\t \mTwo};
\node at (14.35,4.55) {\t 3};
\node at (15.05,4.55) {\t 0};
\node at (15.75,4.55) {\t \mTwo};
\node at (16.45,4.55) {\t \mTwo};
\node at (17.15,4.55) {\t 3};
\node at (17.85,4.55) {\t 0};
\node at (18.55,4.55) {\t \mTwo};
\node at (19.25,4.55) {\t 4};
\node at (19.95,4.55) {\t \mThree};
\node at (20.65,4.55) {\t 0};
\node at (21.35,4.55) {\t 1};
\node at (22.05,4.55) {\t 0};
% row 11
\node at (12.95,3.85) {\t 0};
\node at (13.65,3.85) {\t \mOne};
\node at (14.35,3.85) {\t \mTwo};
\node at (15.05,3.85) {\t 6};
\node at (15.75,3.85) {\t 0};
\node at (16.45,3.85) {\t \mOne};
\node at (17.15,3.85) {\t \mTwo};
\node at (17.85,3.85) {\t 1};
\node at (18.55,3.85) {\t 2};
\node at (19.25,3.85) {\t \mThree};
\node at (19.95,3.85) {\t 0};
\node at (20.65,3.85) {\t \mTwo};
\node at (21.35,3.85) {\t 2};
\node at (22.05,3.85) {\t 0};
% row 12
\node at (12.95,3.15) {\t 0};
\node at (13.65,3.15) {\t 0};
\node at (14.35,3.15) {\t \mOne};
\node at (15.05,3.15) {\t \mTwo};
\node at (15.75,3.15) {\t 3};
\node at (16.45,3.15) {\t 0};
\node at (17.15,3.15) {\t 0};
\node at (17.85,3.15) {\t 0};
\node at (18.55,3.15) {\t 0};
\node at (19.25,3.15) {\t 0};
\node at (19.95,3.15) {\t \mTwo};
\node at (20.65,3.15) {\t 2};
\node at (21.35,3.15) {\t 0};
\node at (22.05,3.15) {\t 0};
% row 13
\node at (12.95,2.45) {\t 0};
\node at (13.65,2.45) {\t 0};
\node at (14.35,2.45) {\t 0};
\node at (15.05,2.45) {\t \mOne};
\node at (15.75,2.45) {\t \mTwo};
\node at (16.45,2.45) {\t 0};
\node at (17.15,2.45) {\t 0};
\node at (17.85,2.45) {\t 0};
\node at (18.55,2.45) {\t 0};
\node at (19.25,2.45) {\t 1};
\node at (19.95,2.45) {\t 2};
\node at (20.65,2.45) {\t 0};
\node at (21.35,2.45) {\t 0};
\node at (22.05,2.45) {\t 0};
% row 14
\node at (12.95,1.75) {\t 0};
\node at (13.65,1.75) {\t 0};
\node at (14.35,1.75) {\t 0};
\node at (15.05,1.75) {\t 0};
\node at (15.75,1.75) {\t 0};
\node at (16.45,1.75) {\t 0};
\node at (17.15,1.75) {\t 0};
\node at (17.85,1.75) {\t 0};
\node at (18.55,1.75) {\t 0};
\node at (19.25,1.75) {\t 0};
\node at (19.95,1.75) {\t 0};
\node at (20.65,1.75) {\t 0};
\node at (21.35,1.75) {\t 0};
\node at (22.05,1.75) {\t 0};

% grid
%\draw[step=0.7,black,thin] (23.79,2.09) grid (33.6,11.2);
\draw[step=0.7,black,thin] (23.79,6.29) grid (28.7,11.2);
\draw[black,ultra thick] (23.8,6.3) rectangle (28.7,11.2);

% max pool value
% row 1
\node at (24.15,10.85) {\t 0};
\node at (24.85,10.85) {\t 2};
\node at (25.55,10.85) {\t 1};
\node at (26.25,10.85) {\t 0};
\node at (26.95,10.85) {\t 0};
\node at (27.65,10.85) {\t 0};
\node at (28.35,10.85) {\t 0};
% row 2
\node at (24.15,10.15) {\t 2};
\node at (24.85,10.15) {\t 2};
\node at (25.55,10.15) {\t 0};
\node at (26.25,10.15) {\t 0};
\node at (26.95,10.15) {\t 3};
\node at (27.65,10.15) {\t 6};
\node at (28.35,10.15) {\t 0};
% row 3
\node at (24.15,9.45) {\t 1};
\node at (24.85,9.45) {\t 0};
\node at (25.55,9.45) {\t 4};
\node at (26.25,9.45) {\t 2};
\node at (26.95,9.45) {\t 2};
\node at (27.65,9.45) {\t 3};
\node at (28.35,9.45) {\t 0};
% row 4
\node at (24.15,8.75) {\t 0};
\node at (24.85,8.75) {\t 2};
\node at (25.55,8.75) {\t 4};
\node at (26.25,8.75) {\t 2};
\node at (26.95,8.75) {\t 1};
\node at (27.65,8.75) {\t 0};
\node at (28.35,8.75) {\t 0};
% row 5
\node at (24.15,8.05) {\t 0};
\node at (24.85,8.05) {\t 3};
\node at (25.55,8.05) {\t 6};
\node at (26.25,8.05) {\t 3};
\node at (26.95,8.05) {\t 4};
\node at (27.65,8.05) {\t 2};
\node at (28.35,8.05) {\t 1};
% row 6
\node at (24.15,7.35) {\t 0};
\node at (24.85,7.35) {\t 6};
\node at (25.55,7.35) {\t 3};
\node at (26.25,7.35) {\t 1};
\node at (26.95,7.35) {\t 2};
\node at (27.65,7.35) {\t 2};
\node at (28.35,7.35) {\t 2};
% row 7
\node at (24.15,6.65) {\t 0};
\node at (24.85,6.65) {\t 0};
\node at (25.55,6.65) {\t 0};
\node at (26.25,6.65) {\t 0};
\node at (26.95,6.65) {\t 1};
\node at (27.65,6.65) {\t 2};
\node at (28.35,6.65) {\t 0};

% line
%\draw[smooth,tension=0.0,thick,->] (1.05,11.2) -- (1.55,11.9) -- (2.05,12.2) -- (11.95,12.2)  -- (12.55,11.9) -- (12.95,11.2);
\draw[smooth,tension=0.0,thick,->] (13.3,11.2) -- (13.8,11.9) -- (14.35,12.2) -- (23.15,12.2)  -- (23.75,11.9) -- (24.15,11.2);

% maxpool
\draw[ultra thick] (12.6,9.8) rectangle (14.0,11.2);

% text
\node at (18.75,12.6) {max};

%% convolution 1
%\draw[ultra thick] (0.0,9.1) rectangle (2.1,11.2);

%% line
%\draw[smooth,tension=0.0,thick,->] (1.05,11.2) -- (1.55,11.9) -- (2.05,12.2) -- (11.95,12.2)  -- (12.55,11.9) -- (12.95,11.2);

%% text
%\node at (7,12.6) {Filter from Figure~\ref{fig:filters}~(a)};

\end{tikzpicture}

%% file: figures/elm.tex
%\begin{tikzpicture}[scale=0.9]
\begin{tikzpicture}[scale=0.75, every node/.style={scale=0.9}]

    % squares (top)
    \draw[thick,color=blue] (1.25,5.5) rectangle (2.25,6.5);
    \draw[thick,color=blue] (3.75,5.5) rectangle (4.75,6.5);
    \draw[thick,color=blue] (6.25,5.5) rectangle (7.25,6.5);
    \draw[thick,color=blue] (8.75,5.5) rectangle (9.75,6.5);
    \draw[thick,color=blue] (11.25,5.5) rectangle (12.25,6.5);
    \node at (14.25,6.0) {$\cdots$};
    \draw[thick,color=blue] (16.25,5.5) rectangle (17.25,6.5);

    \node at (1.75,6.0) {1};
    \node at (4.25,6.0) {2};
    \node at (6.75,6.0) {3};
    \node at (9.25,6.0) {4};
    \node at (11.75,6.0) {5};
    \node at (16.75,6.0) {$n$};
    
    % squares (middle)
    \draw[thick,color=blue] (3.0,2.5) rectangle (4.0,3.5);
    \draw[thick,color=blue] (5.5,2.5) rectangle (6.5,3.5);
    \draw[thick,color=blue] (8.0,2.5) rectangle (9.0,3.5);
    \node at (11.0,3.0) {$\cdots$};
    \draw[thick,color=blue] (13.0,2.5) rectangle (14.0,3.5);

    \node at (3.5,3.0) {1};
    \node at (6.0,3.0) {2};
    \node at (8.5,3.0) {3};
    \node at (13.5,3.0) {$\ell$};
    
    % squares (bottom)
    \draw[thick,color=blue] (4.0,-0.5) rectangle (5.0,0.5);
    \draw[thick,color=blue] (6.5,-0.5) rectangle (7.5,0.5);
    \node at (9.5,0.0) {$\cdots$};
    \draw[thick,color=blue] (12.0,-0.5) rectangle (13.0,0.5);

    \node at (4.5,0.0) {1};
    \node at (7.0,0.0) {2};
    \node at (12.5,0.0) {$m$};

    % top to middle
    \draw[thick,color=blue,->] (1.75,5.5) -- (3.5,3.5);
    \draw[thick,color=blue,->] (1.75,5.5) -- (6.0,3.5);
    \draw[thick,color=blue,->] (1.75,5.5) -- (8.5,3.5);
    \draw[thick,color=blue,->] (1.75,5.5) -- (13.5,3.5);

    \draw[thick,color=blue,->] (4.25,5.5) -- (3.5,3.5);
    \draw[thick,color=blue,->] (4.25,5.5) -- (6.0,3.5);
    \draw[thick,color=blue,->] (4.25,5.5) -- (8.5,3.5);
    \draw[thick,color=blue,->] (4.25,5.5) -- (13.5,3.5);

    \draw[thick,color=blue,->] (6.75,5.5) -- (3.5,3.5);
    \draw[thick,color=blue,->] (6.75,5.5) -- (6.0,3.5);
    \draw[thick,color=blue,->] (6.75,5.5) -- (8.5,3.5);
    \draw[thick,color=blue,->] (6.75,5.5) -- (13.5,3.5);

    \draw[thick,color=blue,->] (9.25,5.5) -- (3.5,3.5);
    \draw[thick,color=blue,->] (9.25,5.5) -- (6.0,3.5);
    \draw[thick,color=blue,->] (9.25,5.5) -- (8.5,3.5);
    \draw[thick,color=blue,->] (9.25,5.5) -- (13.5,3.5);

    \draw[thick,color=blue,->] (11.75,5.5) -- (3.5,3.5);
    \draw[thick,color=blue,->] (11.75,5.5) -- (6.0,3.5);
    \draw[thick,color=blue,->] (11.75,5.5) -- (8.5,3.5);
    \draw[thick,color=blue,->] (11.75,5.5) -- (13.5,3.5);

    \draw[thick,color=blue,->] (16.75,5.5) -- (3.5,3.5);
    \draw[thick,color=blue,->] (16.75,5.5) -- (6.0,3.5);
    \draw[thick,color=blue,->] (16.75,5.5) -- (8.5,3.5);
    \draw[thick,color=blue,->] (16.75,5.5) -- (13.5,3.5);

    % middle to bottom
    \draw[thick,color=blue,->] (3.5,2.5) -- (4.5,0.5);
    \draw[thick,color=blue,->] (3.5,2.5) -- (7.0,0.5);
    \draw[thick,color=blue,->] (3.5,2.5) -- (12.5,0.5);

    \draw[thick,color=blue,->] (6.0,2.5) -- (4.5,0.5);
    \draw[thick,color=blue,->] (6.0,2.5) -- (7.0,0.5);
    \draw[thick,color=blue,->] (6.0,2.5) -- (12.5,0.5);

    \draw[thick,color=blue,->] (8.5,2.5) -- (4.5,0.5);
    \draw[thick,color=blue,->] (8.5,2.5) -- (7.0,0.5);
    \draw[thick,color=blue,->] (8.5,2.5) -- (12.5,0.5);

    \draw[thick,color=blue,->] (13.5,2.5) -- (4.5,0.5);
    \draw[thick,color=blue,->] (13.5,2.5) -- (7.0,0.5);
    \draw[thick,color=blue,->] (13.5,2.5) -- (12.5,0.5);
    
    % labels
%    \node at (18.0,6.0) {$X$};
%    \node at (16.125,4.25) {$W$};
%    \node at (14.75,3.0) {$H$};
%    \node at (13.75,1.5) {$\beta$};
%    \node at (13.75,0.0) {$Y$};
    \node at (0.65,6.0) {$X$};
    \node at (1.75,4.25) {$W$};
    \node at (2.4,3.0) {$H$};
    \node at (3.25,1.5) {$\beta$};
    \node at (3.4,0.0) {$Y$};

\end{tikzpicture}

%% file: figures/fig_cnn_c1.tex
%\begin{tikzpicture}[scale=0.9, every node/.style={scale=1.0}, rotate=-90]
\begin{tikzpicture}[scale=0.9, every node/.style={scale=1.0}]
    \begin{axis}[
        width  = 1.0*\textwidth,
        height = 10cm,
        ymin=0.0,ymax=1.095,
        ytick={0.0,0.2,0.4,0.6,0.8,1.0},
        major x tick style = transparent,
%        ybar=5.0*\pgflinewidth,
        ybar=3.75*\pgflinewidth,
%        ybar,
%        bar width=7.0pt,
        bar width=5.5pt,
%        ymajorgrids = true,
        xlabel = {Malware family},
        ylabel = {Accuracy},
        symbolic x coords={Adialer.C,Agent.FYI,Allaple.A,Allaple.L,Alueron.gen!J,Autorun.K,C2LOP.P,C2LOP.gen!g,Dialplatform.B,Dontovo.A,Fakerean,Instantaccess,Lolyda.AA1,Lolyda.AA2,Lolyda.AA3,Lolyda.AT,Malex.gen!J,Obfuscator.AD,Rbot!gen,Skintrim.N,Swizzor.gen!E,Swizzor.gen!I,VB.AT,Wintrim.BX,Yuner.A},
	y tick label style={
    		/pgf/number format/.cd,
   		fixed,
   		fixed zerofill,
    		precision=1},
%	yticklabel pos=right,
        xtick = data,
        x tick label style={
        	        rotate=60,
		font=\footnotesize\tt,
		anchor=north east,
		inner sep=0mm
		},
%		font=\small},
%        scaled y ticks = false,
	%%%%% numbers on bars and rotated
        nodes near coords,
        every node near coord/.append style={/pgf/number format/precision=2,
        								  rotate=90, 
								  scale=0.80,
        								  anchor=west,
								  font=\footnotesize},
        %%%%%
        enlarge x limits=0.025,
        legend cell align=left,
        legend style={
%                at={(1,1.05)},
%                anchor=south east,
%	        nodes={rotate=90},%%%%% rotate text in legend
%                at={(0.125,0)},
%                at={(0.125,0)},
%                at={(0.8775,0)},
%                at={(0.135,0.015)},
                at={(0.905,0.015)},
                anchor=south,
                column sep=1ex
        }
    ]
\addplot[fill=blue,opacity=1.00] %%%%% 32 filters
coordinates {
(Adialer.C,1.00)
(Agent.FYI,1.00)
(Allaple.A,1.00)
(Allaple.L,1.00)
(Alueron.gen!J,0.89)
(Autorun.K,0.00)
(C2LOP.P,0.62)
(C2LOP.gen!g,0.79)
(Dialplatform.B,1.00)
(Dontovo.A,1.00)
(Fakerean,1.00)
(Instantaccess,1.00)
(Lolyda.AA1,1.00)
(Lolyda.AA2,1.00)
(Lolyda.AA3,0.82)
(Lolyda.AT,0.93)
(Malex.gen!J,0.92)
(Obfuscator.AD,1.00)
(Rbot!gen,0.93)
(Skintrim.N,1.00)
(Swizzor.gen!E,0.83)
(Swizzor.gen!I,0.58)
(VB.AT,1.00)
(Wintrim.BX,0.88)
(Yuner.A,1.00)
};
\addplot[fill=red,opacity=1.00] %%%%% 64 filters
coordinates {
(Adialer.C,1.00)
(Agent.FYI,1.00)
(Allaple.A,1.00)
(Allaple.L,1.00)
(Alueron.gen!J,1.00)
(Autorun.K,0.00)
(C2LOP.P,0.46)
(C2LOP.gen!g,1.00)
(Dialplatform.B,1.00)
(Dontovo.A,1.00)
(Fakerean,1.00)
(Instantaccess,1.00)
(Lolyda.AA1,1.00)
(Lolyda.AA2,1.00)
(Lolyda.AA3,0.82)
(Lolyda.AT,0.93)
(Malex.gen!J,0.42)
(Obfuscator.AD,1.00)
(Rbot!gen,1.00)
(Skintrim.N,1.00)
(Swizzor.gen!E,0.75)
(Swizzor.gen!I,0.50)
(VB.AT,0.98)
(Wintrim.BX,0.88)
(Yuner.A,1.00)
};
\legend{32 filters,64 filters}
\end{axis}
\end{tikzpicture}

%% file: figures/conf_exp2.tex
%\begin{tikzpicture}[scale=0.4]
\begin{tikzpicture}[scale=0.6]
    \begin{axis}[%colorbar/width=2.5mm,
        width=18cm,
        height=18cm,
%        colormap={blackwhite}{gray(0cm)=(1); gray(1cm)=(0.5)},
%	colormap={bluewhite}{color=(white) color=(blue)},
%	colormap={bluewhite}{color=(white) rgb255=(0,191,255)},
	colormap={bluewhite}{color=(white) rgb255=(100,149,237)},
        xticklabels={Adialer.C,Agent.FYI,Allaple.A,Allaple.L,Alueron.gen!J,Autorun.K,C2LOP.P,C2LOP.gen!g,Dialplatform.B,Dontovo.A,Fakerean,Instantaccess,Lolyda.AA1,Lolyda.AA2,Lolyda.AA3,Lolyda.AT,Malex.gen!J,Obfuscator.AD,Rbot!gen,Skintrim.N,Swizzor.gen!E,Swizzor.gen!I,VB.AT,Wintrim.BX,Yuner.A},
        xtick={0,...,25},
        xtick style={draw=none},
	xticklabel style={anchor=east,rotate=45,yshift=-5pt},
        yticklabels={Adialer.C,Agent.FYI,Allaple.A,Allaple.L,Alueron.gen!J,Autorun.K,C2LOP.P,C2LOP.gen!g,Dialplatform.B,Dontovo.A,Fakerean,Instantaccess,Lolyda.AA1,Lolyda.AA2,Lolyda.AA3,Lolyda.AT,Malex.gen!J,Obfuscator.AD,Rbot!gen,Skintrim.N,Swizzor.gen!E,Swizzor.gen!I,VB.AT,Wintrim.BX,Yuner.A},
        ytick={0,...,24},
        ytick style={draw=none},
        enlargelimits=false,
        xlabel style={font=\footnotesize},
        ylabel style={font=\footnotesize},
        legend style={font=\footnotesize},
        xticklabel style={font=\footnotesize\tt},
        yticklabel style={font=\footnotesize\tt},
        colorbar,
        colorbar style={
%     	  	width=0.05*\pgfkeysvalueof{/pgfplots/parent axis width},%%% added this
%     	  	height=0.5*\pgfkeysvalueof{/pgfplots/parent axis height},
%		plot graphics/node/.style={scale=1.33,anchor=south west,inner sep=0pt,}, %%% scale colorbar fill %%%
            ytick={0,0.20,0.40,0.60,0.80,1.00},
            yticklabels={0,0.20,0.40,0.60,0.80,1.00},
            yticklabel={\pgfmathprintnumber\tick},
            yticklabel style={font=\footnotesize,
            		/pgf/number format/fixed,
			/pgf/number format/precision=1}
        },
%        point meta min=0,
%        point meta max=100,
        point meta min=0.0,
        point meta max=1.0,
        nodes near coords={\pgfmathprintnumber\pgfplotspointmeta},
        % ---------------------------------------------------------------------
        % show `nodes near coords' but adapt the style so that values
        % above a threshold get another style
        % (adapted from <http://tex.stackexchange.com/a/141006/95441>)
        % #1: the THRESHOLD after which we switch to a special display.
        nodes near coords black white/.style={
            % define the style of the nodes with "small" values
            small value/.style={
                font=\scriptsize,
                yshift=-7pt,
%                text=white,
                text=black,
                /pgf/number format/fixed,
                /pgf/number format/precision=2
%                /pgf/number format/precision=0
            },
            % define the style of the nodes with "large" values
            large value/.style={
                font=\scriptsize,
                yshift=-7pt,
%                text=black,
                text=white,
                /pgf/number format/fixed,
                /pgf/number format/precision=2
%                /pgf/number format/precision=0
            },
            every node near coord/.style={
                check for zero/.code={
                    \pgfmathfloatifflags{\pgfplotspointmeta}{0}{
                        % If meta=0, make the node a coordinate
                        % (which doesn't have text)
                        \pgfkeys{/tikz/coordinate}
                    }{
                        \begingroup
                        % this group is merely to switch to FPU locally.
                        % Might be unnecessary, but who knows.
                        \pgfkeys{/pgf/fpu}
                        \pgfmathparse{\pgfplotspointmeta<#1}
                        \global\let\result=\pgfmathresult
                        \endgroup
                        %
                        % simplifies debugging:
                        %\show\result
                        %
                        \pgfmathfloatcreate{1}{1.0}{0}
                        \let\ONE=\pgfmathresult
                        \ifx\result\ONE
                            % AH: our condition 'y < #1' is met.
                            \pgfkeysalso{/pgfplots/small value}
                        \else
                            % ok, proceed as usual.
                            \pgfkeysalso{/pgfplots/large value}
                        \fi
                    }
                },
                check for zero,
            },
        },
        % asign a value to the new style which is the threshold at which
        % the two style `small value' or `large value' are used
%        nodes near coords black white=50,
        nodes near coords black white=0.5,
        % -----------------------------------------------------------------
    ]
        \addplot[
            matrix plot,
            mesh/cols=25,
            point meta=explicit,draw=gray
        ] table [meta=C] {
            x y C
 0  0 1.00
 1  0 0.00
 2  0 0.00
 3  0 0.00
 4  0 0.00
 5  0 0.00
 6  0 0.00
 7  0 0.00
 8  0 0.00
 9  0 0.00
10  0 0.00
11  0 0.00
12  0 0.00
13  0 0.00
14  0 0.00
15  0 0.00
16  0 0.00
17  0 0.00
18  0 0.00
19  0 0.00
20  0 0.00
21  0 0.00
22  0 0.00
23  0 0.00
24  0 0.00
 0  1 0.00
 1  1 1.00
 2  1 0.00
 3  1 0.00
 4  1 0.00
 5  1 0.00
 6  1 0.00
 7  1 0.00
 8  1 0.00
 9  1 0.00
10  1 0.00
11  1 0.00
12  1 0.00
13  1 0.00
14  1 0.00
15  1 0.00
16  1 0.00
17  1 0.00
18  1 0.00
19  1 0.00
20  1 0.00
21  1 0.00
22  1 0.00
23  1 0.00
24  1 0.00
 0  2 0.00
 1  2 0.00
 2  2 1.00
 3  2 0.00
 4  2 0.00
 5  2 0.00
 6  2 0.00
 7  2 0.00
 8  2 0.00
 9  2 0.00
10  2 0.00
11  2 0.00
12  2 0.00
13  2 0.00
14  2 0.00
15  2 0.00
16  2 0.00
17  2 0.00
18  2 0.00
19  2 0.00
20  2 0.00
21  2 0.00
22  2 0.00
23  2 0.00
24  2 0.00
 0  3 0.00
 1  3 0.00
 2  3 0.00
 3  3 1.00
 4  3 0.00
 5  3 0.00
 6  3 0.00
 7  3 0.00
 8  3 0.00
 9  3 0.00
10  3 0.00
11  3 0.00
12  3 0.00
13  3 0.00
14  3 0.00
15  3 0.00
16  3 0.00
17  3 0.00
18  3 0.00
19  3 0.00
20  3 0.00
21  3 0.00
22  3 0.00
23  3 0.00
24  3 0.00
 0  4 0.00
 1  4 0.00
 2  4 0.00
 3  4 0.00
 4  4 0.89
 5  4 0.00
 6  4 0.00
 7  4 0.00
 8  4 0.00
 9  4 0.00
10  4 0.00
11  4 0.00
12  4 0.00
13  4 0.00
14  4 0.00
15  4 0.00
16  4 0.00
17  4 0.11
18  4 0.00
19  4 0.00
20  4 0.00
21  4 0.00
22  4 0.00
23  4 0.00
24  4 0.00
 0  5 0.00
 1  5 0.00
 2  5 0.00
 3  5 0.00
 4  5 0.00
 5  5 0.00
 6  5 0.00
 7  5 0.00
 8  5 0.00
 9  5 0.00
10  5 0.00
11  5 0.00
12  5 0.00
13  5 0.00
14  5 0.00
15  5 0.00
16  5 0.00
17  5 0.00
18  5 0.00
19  5 0.00
20  5 0.00
21  5 0.00
22  5 0.00
23  5 0.00
24  5 1.00
 0  6 0.00
 1  6 0.00
 2  6 0.00
 3  6 0.00
 4  6 0.00
 5  6 0.00
 6  6 0.62
 7  6 0.38
 8  6 0.00
 9  6 0.00
10  6 0.00
11  6 0.00
12  6 0.00
13  6 0.00
14  6 0.00
15  6 0.00
16  6 0.00
17  6 0.00
18  6 0.00
19  6 0.00
20  6 0.00
21  6 0.00
22  6 0.00
23  6 0.00
24  6 0.00
 0  7 0.00
 1  7 0.00
 2  7 0.00
 3  7 0.00
 4  7 0.00
 5  7 0.00
 6  7 0.00
 7  7 0.79
 8  7 0.00
 9  7 0.00
10  7 0.00
11  7 0.00
12  7 0.00
13  7 0.00
14  7 0.00
15  7 0.00
16  7 0.00
17  7 0.00
18  7 0.00
19  7 0.00
20  7 0.16
21  7 0.05
22  7 0.00
23  7 0.00
24  7 0.00
 0  8 0.00
 1  8 0.00
 2  8 0.00
 3  8 0.00
 4  8 0.00
 5  8 0.00
 6  8 0.00
 7  8 0.00
 8  8 1.00
 9  8 0.00
10  8 0.00
11  8 0.00
12  8 0.00
13  8 0.00
14  8 0.00
15  8 0.00
16  8 0.00
17  8 0.00
18  8 0.00
19  8 0.00
20  8 0.00
21  8 0.00
22  8 0.00
23  8 0.00
24  8 0.00
 0  9 0.00
 1  9 0.00
 2  9 0.00
 3  9 0.00
 4  9 0.00
 5  9 0.00
 6  9 0.00
 7  9 0.00
 8  9 0.00
 9  9 1.00
10  9 0.00
11  9 0.00
12  9 0.00
13  9 0.00
14  9 0.00
15  9 0.00
16  9 0.00
17  9 0.00
18  9 0.00
19  9 0.00
20  9 0.00
21  9 0.00
22  9 0.00
23  9 0.00
24  9 0.00
 0 10 0.00
 1 10 0.00
 2 10 0.00
 3 10 0.00
 4 10 0.00
 5 10 0.00
 6 10 0.00
 7 10 0.00
 8 10 0.00
 9 10 0.00
10 10 1.00
11 10 0.00
12 10 0.00
13 10 0.00
14 10 0.00
15 10 0.00
16 10 0.00
17 10 0.00
18 10 0.00
19 10 0.00
20 10 0.00
21 10 0.00
22 10 0.00
23 10 0.00
24 10 0.00
 0 11 0.00
 1 11 0.00
 2 11 0.00
 3 11 0.00
 4 11 0.00
 5 11 0.00
 6 11 0.00
 7 11 0.00
 8 11 0.00
 9 11 0.00
10 11 0.00
11 11 1.00
12 11 0.00
13 11 0.00
14 11 0.00
15 11 0.00
16 11 0.00
17 11 0.00
18 11 0.00
19 11 0.00
20 11 0.00
21 11 0.00
22 11 0.00
23 11 0.00
24 11 0.00
 0 12 0.00
 1 12 0.00
 2 12 0.00
 3 12 0.00
 4 12 0.00
 5 12 0.00
 6 12 0.00
 7 12 0.00
 8 12 0.00
 9 12 0.00
10 12 0.00
11 12 0.00
12 12 1.00
13 12 0.00
14 12 0.00
15 12 0.00
16 12 0.00
17 12 0.00
18 12 0.00
19 12 0.00
20 12 0.00
21 12 0.00
22 12 0.00
23 12 0.00
24 12 0.00
 0 13 0.00
 1 13 0.00
 2 13 0.00
 3 13 0.00
 4 13 0.00
 5 13 0.00
 6 13 0.00
 7 13 0.00
 8 13 0.00
 9 13 0.00
10 13 0.00
11 13 0.00
12 13 0.00
13 13 1.00
14 13 0.00
15 13 0.00
16 13 0.00
17 13 0.00
18 13 0.00
19 13 0.00
20 13 0.00
21 13 0.00
22 13 0.00
23 13 0.00
24 13 0.00
 0 14 0.00
 1 14 0.00
 2 14 0.00
 3 14 0.00
 4 14 0.00
 5 14 0.00
 6 14 0.00
 7 14 0.00
 8 14 0.00
 9 14 0.00
10 14 0.00
11 14 0.00
12 14 0.00
13 14 0.00
14 14 0.82
15 14 0.00
16 14 0.00
17 14 0.00
18 14 0.00
19 14 0.00
20 14 0.00
21 14 0.00
22 14 0.18
23 14 0.00
24 14 0.00
 0 15 0.00
 1 15 0.00
 2 15 0.00
 3 15 0.00
 4 15 0.00
 5 15 0.00
 6 15 0.00
 7 15 0.00
 8 15 0.00
 9 15 0.00
10 15 0.00
11 15 0.00
12 15 0.00
13 15 0.00
14 15 0.00
15 15 0.93
16 15 0.00
17 15 0.00
18 15 0.07
19 15 0.00
20 15 0.00
21 15 0.00
22 15 0.00
23 15 0.00
24 15 0.00
 0 16 0.00
 1 16 0.00
 2 16 0.00
 3 16 0.00
 4 16 0.00
 5 16 0.00
 6 16 0.00
 7 16 0.00
 8 16 0.00
 9 16 0.00
10 16 0.00
11 16 0.00
12 16 0.00
13 16 0.08
14 16 0.00
15 16 0.00
16 16 0.92
17 16 0.00
18 16 0.00
19 16 0.00
20 16 0.00
21 16 0.00
22 16 0.00
23 16 0.00
24 16 0.00
 0 17 0.00
 1 17 0.00
 2 17 0.00
 3 17 0.00
 4 17 0.00
 5 17 0.00
 6 17 0.00
 7 17 0.00
 8 17 0.00
 9 17 0.00
10 17 0.00
11 17 0.00
12 17 0.00
13 17 0.00
14 17 0.00
15 17 0.00
16 17 0.00
17 17 1.00
18 17 0.00
19 17 0.00
20 17 0.00
21 17 0.00
22 17 0.00
23 17 0.00
24 17 0.00
 0 18 0.00
 1 18 0.00
 2 18 0.00
 3 18 0.00
 4 18 0.00
 5 18 0.00
 6 18 0.00
 7 18 0.00
 8 18 0.00
 9 18 0.00
10 18 0.00
11 18 0.00
12 18 0.00
13 18 0.00
14 18 0.00
15 18 0.00
16 18 0.00
17 18 0.00
18 18 0.93
19 18 0.00
20 18 0.00
21 18 0.07
22 18 0.00
23 18 0.00
24 18 0.00
 0 19 0.00
 1 19 0.00
 2 19 0.00
 3 19 0.00
 4 19 0.00
 5 19 0.00
 6 19 0.00
 7 19 0.00
 8 19 0.00
 9 19 0.00
10 19 0.00
11 19 0.00
12 19 0.00
13 19 0.00
14 19 0.00
15 19 0.00
16 19 0.00
17 19 0.00
18 19 0.00
19 19 1.00
20 19 0.00
21 19 0.00
22 19 0.00
23 19 0.00
24 19 0.00
 0 20 0.00
 1 20 0.00
 2 20 0.00
 3 20 0.00
 4 20 0.00
 5 20 0.00
 6 20 0.00
 7 20 0.00
 8 20 0.00
 9 20 0.00
10 20 0.00
11 20 0.00
12 20 0.00
13 20 0.00
14 20 0.00
15 20 0.00
16 20 0.00
17 20 0.00
18 20 0.00
19 20 0.00
20 20 0.83
21 20 0.17
22 20 0.00
23 20 0.00
24 20 0.00
 0 21 0.00
 1 21 0.00
 2 21 0.00
 3 21 0.00
 4 21 0.00
 5 21 0.00
 6 21 0.00
 7 21 0.00
 8 21 0.00
 9 21 0.00
10 21 0.00
11 21 0.00
12 21 0.00
13 21 0.00
14 21 0.00
15 21 0.00
16 21 0.00
17 21 0.00
18 21 0.00
19 21 0.00
20 21 0.42
21 21 0.58
22 21 0.00
23 21 0.00
24 21 0.00
 0 22 0.00
 1 22 0.00
 2 22 0.00
 3 22 0.00
 4 22 0.00
 5 22 0.00
 6 22 0.00
 7 22 0.00
 8 22 0.00
 9 22 0.00
10 22 0.00
11 22 0.00
12 22 0.00
13 22 0.00
14 22 0.00
15 22 0.00
16 22 0.00
17 22 0.00
18 22 0.00
19 22 0.00
20 22 0.00
21 22 0.00
22 22 1.00
23 22 0.00
24 22 0.00
 0 23 0.00
 1 23 0.00
 2 23 0.00
 3 23 0.00
 4 23 0.00
 5 23 0.00
 6 23 0.12
 7 23 0.00
 8 23 0.00
 9 23 0.00
10 23 0.00
11 23 0.00
12 23 0.00
13 23 0.00
14 23 0.00
15 23 0.00
16 23 0.00
17 23 0.00
18 23 0.00
19 23 0.00
20 23 0.00
21 23 0.00
22 23 0.00
23 23 0.88
24 23 0.00
 0 24 0.00
 1 24 0.00
 2 24 0.00
 3 24 0.00
 4 24 0.00
 5 24 0.00
 6 24 0.00
 7 24 0.00
 8 24 0.00
 9 24 0.00
10 24 0.00
11 24 0.00
12 24 0.00
13 24 0.00
14 24 0.00
15 24 0.00
16 24 0.00
17 24 0.00
18 24 0.00
19 24 0.00
20 24 0.00
21 24 0.00
22 24 0.00
23 24 0.00
24 24 1.00
        };
    \end{axis}
\end{tikzpicture}
%
%\caption{I'm confused~5!}\label{tab:CM5}
%\end{figure*}

%% file: figures/fig_cnn_c2.tex
%\begin{tikzpicture}[scale=0.9, every node/.style={scale=1.0}, rotate=-90]
\begin{tikzpicture}[scale=0.9, every node/.style={scale=1.0}]
    \begin{axis}[
        width  = 0.85*\textwidth,
        height = 9.0cm,
        ymin=0.0,ymax=1.12,
        ytick={0.0,0.2,0.4,0.6,0.8,1.0},
        major x tick style = transparent,
%        ybar=5.0*\pgflinewidth,
%        ybar=3.75*\pgflinewidth,
        ybar,
        bar width=8.0pt,
%        bar width=5.5pt,
%        ymajorgrids = true,
        xlabel = {Malware family},
        ylabel = {Accuracy},
        symbolic x coords={Adialer.C,Agent.FYI,Allaple.A,Allaple.L,Alueron.gen!J,Autorun.K,C2LOP.P,C2LOP.gen!g,Dialplatform.B,Dontovo.A,Fakerean,Instantaccess,Lolyda.AA1,Lolyda.AA2,Lolyda.AA3,Lolyda.AT,Malex.gen!J,Obfuscator.AD,Rbot!gen,Skintrim.N,Swizzor.gen!E,Swizzor.gen!I,VB.AT,Wintrim.BX,Yuner.A},
	y tick label style={
    		/pgf/number format/.cd,
   		fixed,
   		fixed zerofill,
    		precision=1},
%	yticklabel pos=right,
        xtick = data,
        x tick label style={
        	        rotate=60,
		font=\footnotesize\tt,
		anchor=north east,
		inner sep=0mm
		},
%		font=\small},
%        scaled y ticks = false,
	%%%%% numbers on bars and rotated
        nodes near coords,
        every node near coord/.append style={/pgf/number format/precision=2,
        								  rotate=90, 
								  scale=0.90,
        								  anchor=west,
								  font=\footnotesize},
        %%%%%
        enlarge x limits=0.05,
        legend cell align=left,
        legend style={
%                at={(1,1.05)},
%                anchor=south east,
%	        nodes={rotate=90},%%%%% rotate text in legend
%                at={(0.125,0)},
%                at={(0.125,0)},
%                at={(0.8775,0)},
%                at={(0.135,0.015)},
                at={(0.905,0.015)},
                anchor=south,
                column sep=1ex
        }
    ]
\addplot[fill=blue,opacity=1.00]
coordinates {
(Adialer.C,1.00)
(Agent.FYI,1.00)
(Allaple.A,1.00)
(Allaple.L,1.00)
(Alueron.gen!J,1.00)
(Autorun.K,0.00)
(C2LOP.P,0.38)
(C2LOP.gen!g,0.74)
(Dialplatform.B,1.00)
(Dontovo.A,1.00)
(Fakerean,1.00)
(Instantaccess,1.00)
(Lolyda.AA1,0.95)
(Lolyda.AA2,1.00)
(Lolyda.AA3,0.82)
(Lolyda.AT,0.93)
(Malex.gen!J,0.92)
(Obfuscator.AD,1.00)
(Rbot!gen,0.71)
(Skintrim.N,1.00)
(Swizzor.gen!E,0.92)
(Swizzor.gen!I,0.50)
(VB.AT,1.00)
(Wintrim.BX,0.88)
(Yuner.A,1.00)
};
%\legend{32 filters,64 filters}
\end{axis}
\end{tikzpicture}

%% file: figures/fig_act_elm_1.tex
\begin{tikzpicture}[scale=0.8, every node/.style={scale=1.0}]
    \begin{axis}[
        width  = 0.95*\textwidth,
        height = 10cm,
        ymin=0.0,ymax=1.0475,
        ytick={0.0,0.2,0.4,0.6,0.8,1.0},
        major x tick style = transparent,
        ybar=5.0*\pgflinewidth,
        bar width=8.0pt,
%        ymajorgrids = true,
        xlabel = {Number of neurons},
        ylabel = {Average testing accuracy},
        symbolic x coords={128,
        				      256,
				      512,
				      1024,
				      2048,
				      4096},
	y tick label style={
    		/pgf/number format/.cd,
   		fixed,
   		fixed zerofill,
    		precision=1},
%	yticklabel pos=right,
        xtick = data,
        x tick label style={
%        	        rotate=60,
%		font=\footnotesize,
%		anchor=north east,
%		inner sep=0mm
		},
%		font=\small},
%        scaled y ticks = false,
	%%%%% numbers on bars and rotated
        nodes near coords,
        every node near coord/.append style={/pgf/number format/precision=3,
        								  rotate=90, 
        								  anchor=west,
								  font=\footnotesize},
        %%%%%
        enlarge x limits=0.11,
        legend cell align=left,
        legend style={
%                at={(1,1.05)},
%                anchor=south east,
%	        nodes={rotate=90},%%%%% rotate text in legend
%                at={(0.125,0)},
%                at={(0.125,0)},
%                at={(0.8775,0)},
%                at={(0.135,0.015)},
                at={(0.875,0.015)},
                anchor=south,
                column sep=1ex
        }
    ]
\addplot[fill=green,opacity=1.00] %%%%% tanh
coordinates {
(128,0.659)
(256,0.737)
(512,0.801)
(1024,0.844)
(2048,0.856)
(4096,0.852)
};
\addplot[fill=black,opacity=1.00] %%%%% relu
coordinates {
(128,0.699)
(256,0.775)
(512,0.827)
(1024,0.874)
(2048,0.899)
(4096,0.892)
};
\addplot[fill=red,opacity=1.00] %%%%% softlim
coordinates {
(128,0.655)
(256,0.739)
(512,0.801)
(1024,0.845)
(2048,0.869)
(4096,0.856)
};
\addplot[fill=blue,opacity=1.00] %%%%% hardlim
coordinates {
(128,0.652)
(256,0.737)
(512,0.799)
(1024,0.842)
(2048,0.869)
(4096,0.855)
};
\addplot[fill=yellow,opacity=1.00] %%%%% multiquadric
coordinates {
(128,0.706)
(256,0.776)
(512,0.827)
(1024,0.872)
(2048,0.901)
(4096,0.893)
};
\legend{tanh,relu,softlim,hardlim,multiquadric}
\end{axis}
\end{tikzpicture}

%% file: figures/elm_1_train_test.tex
%\begin{tikzpicture}[scale=0.75]
\begin{tikzpicture}[scale=0.875]
\begin{axis}[width=0.90\textwidth,
		   height=0.675\textwidth,
%		   /pgf/number format/1000 sep={},
   	           symbolic x coords={128,
	           				 256,
						 512,
						 1024,
						 2048,
						 4096},
	 	   x tick label style={
   		 	/pgf/number format/.cd,
			/pgf/number format/1000 sep={},
   			fixed,
   			fixed zerofill,
    			precision=0
		   },
	 	   y tick label style={
    		 	/pgf/number format/.cd,
   			fixed,
   			fixed zerofill,
    			precision=1
		    },
                    ymin=0.5,ymax=1.0,
                    legend pos=south east,
                    legend cell align={left},
                    xtick={128,256,512,1024,2048,4096},
                    ytick={0.5,0.6,0.7,0.8,0.9,1.0},
                    xlabel={Number of neurons},
                    ylabel={Accuracy}] 
\addplot[color=green,ultra thick,mark=*,mark size=2.0] coordinates { % tanh test
    (128,0.659)
    (256,0.737)
    (512,0.801)
    (1024,0.844)
    (2048,0.856)
    (4096,0.852)
};
\addplot[color=black,ultra thick,mark=*,mark size=2.0] coordinates { % relu test
    (128,0.699)
    (256,0.775)
    (512,0.827)
    (1024,0.874)
    (2048,0.899)
    (4096,0.892)
};
\addplot[color=red,ultra thick,mark=*,mark size=2.0] coordinates { % softlim test
    (128,0.655)
    (256,0.739)
    (512,0.801)
    (1024,0.845)
    (2048,0.869)
    (4096,0.856)
};
\addplot[color=blue,ultra thick,mark=*,mark size=2.0] coordinates { % hardlim test
    (128,0.652)
    (256,0.737)
    (512,0.799)
    (1024,0.842)
    (2048,0.869)
    (4096,0.855)
};
\addplot[color=yellow,ultra thick,mark=*,mark size=2.0] coordinates { % multiquadric test
    (128,0.706)
    (256,0.776)
    (512,0.827)
    (1024,0.872)
    (2048,0.901)
    (4096,0.893)
};
\addplot[color=green,dashed,ultra thick,mark=*,mark size=2.0] coordinates { % tanh train
    (128,0.668)
    (256,0.762)
    (512,0.845)
    (1024,0.915)
    (2048,0.960)
    (4096,1.000)
};
\addplot[color=black,dashed,ultra thick,mark=*,mark size=2.0] coordinates { % relu train
    (128,0.710)
    (256,0.796)
    (512,0.866)
    (1024,0.937)
    (2048,0.987)
    (4096,1.000)
};
\addplot[color=red,dashed,ultra thick,mark=*,mark size=2.0] coordinates { % softlim train
    (128,0.666)
    (256,0.764)
    (512,0.844)
    (1024,0.916)
    (2048,0.975)
    (4096,1.000)
};
\addplot[color=blue,dashed,ultra thick,mark=*,mark size=2.0] coordinates { % hardlim train
    (128,0.663)
    (256,0.763)
    (512,0.846)
    (1024,0.916)
    (2048,0.975)
    (4096,1.000)
};
\addplot[color=yellow,dashed,ultra thick,mark=*,mark size=2.0] coordinates { % multiquadric train
    (128,0.717)
    (256,0.800)
    (512,0.866)
    (1024,0.935)
    (2048,0.989)
    (4096,1.000)
};
\legend{relu test,tanh test,softlim test,hardlim test,multiquadric test,
	relu train,tanh train,softlim train,hardlim train,multiquadric train}
\end{axis}
\end{tikzpicture}

%% file: figures/fig_act_elm_05.tex
\begin{tikzpicture}[scale=0.9, every node/.style={scale=1.0}]
    \begin{axis}[
        width  = 0.95*\textwidth,
        height = 10cm,
        ymin=0.0,ymax=1.0475,
        ytick={0.0,0.2,0.4,0.6,0.8,1.0},
        major x tick style = transparent,
        ybar=5.0*\pgflinewidth,
        bar width=8.0pt,
%        ymajorgrids = true,
        xlabel = {Number of neurons},
        ylabel = {Average testing accuracy},
        symbolic x coords={128,
        				      256,
				      512,
				      1024,
				      2048,
				      4096},
	y tick label style={
    		/pgf/number format/.cd,
   		fixed,
   		fixed zerofill,
    		precision=1},
%	yticklabel pos=right,
        xtick = data,
        x tick label style={
%        	        rotate=60,
%		font=\footnotesize,
%		anchor=north east,
%		inner sep=0mm
		},
%		font=\small},
%        scaled y ticks = false,
	%%%%% numbers on bars and rotated
        nodes near coords,
        every node near coord/.append style={/pgf/number format/precision=3,
        								  rotate=90, 
        								  anchor=west,
								  font=\footnotesize},
        %%%%%
        enlarge x limits=0.11,
        legend cell align=left,
        legend style={
%                at={(1,1.05)},
%                anchor=south east,
%	        nodes={rotate=90},%%%%% rotate text in legend
%                at={(0.125,0)},
%                at={(0.125,0)},
%                at={(0.8775,0)},
%                at={(0.135,0.015)},
                at={(0.875,0.015)},
                anchor=south,
                column sep=1ex
        }
    ]
\addplot[fill=green,opacity=1.00] %%%%% tanh
coordinates {
(128,0.653)
(256,0.738)
(512,0.802)
(1024,0.845)
(2048,0.870)
(4096,0.850)
};
\addplot[fill=black,opacity=1.00] %%%%% relu
coordinates {
(128,0.696)
(256,0.771)
(512,0.827)
(1024,0.872)
(2048,0.900)
(4096,0.894)
};
\addplot[fill=red,opacity=1.00] %%%%% softlim
coordinates {
(128,0.654)
(256,0.736)
(512,0.800)
(1024,0.840)
(2048,0.868)
(4096,0.854)
};
\addplot[fill=blue,opacity=1.00] %%%%% hardlim
coordinates {
(128,0.656)
(256,0.739)
(512,0.798)
(1024,0.842)
(2048,0.869)
(4096,0.853)
};
\addplot[fill=yellow,opacity=1.00] %%%%% multiquadric
coordinates {
(128,0.707)
(256,0.776)
(512,0.828)
(1024,0.873)
(2048,0.902)
(4096,0.893)
};
\legend{tanh,relu,softlim,hardlim,multiquadric}
\end{axis}
\end{tikzpicture}

%% file: figures/fig_elm_train_time.tex
\begin{tikzpicture}[scale=0.9, every node/.style={scale=1.0}]
    \begin{axis}[
        width  = 0.6*\textwidth,
        height = 8cm,
        ymin=0.0,ymax=5300,
        ytick={0,1000,2000,3000,4000,5000},
        major x tick style = transparent,
        ybar=7.0*\pgflinewidth,
        bar width=12.0pt,
%        ymajorgrids = true,
        xlabel = {Number of neurons},
        ylabel = {Time (seconds)},
        symbolic x coords={128,
        				      256,
				      512,
				      1024},
	y tick label style={
    		/pgf/number format/.cd,
		1000 sep={},
   		fixed,
   		fixed zerofill,
    		precision=0},
%	yticklabel pos=right,
        xtick = data,
%        x tick label style={
%        	        rotate=60,
%		font=\footnotesize,
%		anchor=north east,
%		inner sep=0mm
%		},
%		font=\small},
%        scaled y ticks = false,
	%%%%% numbers on bars and rotated
        nodes near coords,
        every node near coord/.append style={/pgf/number format/precision=0,
        								  rotate=90, 
        								  anchor=west,
								  font=\footnotesize,
								  /pgf/number format/1000 sep=},
        %%%%%
        enlarge x limits=0.15,
        legend cell align=left,
        legend style={
%                at={(1,1.05)},
%                anchor=south east,
%	        nodes={rotate=90},%%%%% rotate text in legend
%                at={(0.125,0)},
%                at={(0.125,0)},
%                at={(0.8775,0)},
%                at={(0.135,0.015)},
%                at={(0.82,0.015)},
                at={(0.1525,0.8)},
                anchor=south,
                column sep=1ex
        }
    ]
\addplot[fill=red,opacity=1.00] %%%%% average case
coordinates {
(128,49.74)
(256,78.18)
(512,128.44)
(1024,294.42)
};
\addplot[fill=blue,opacity=1.00] %%%%% ensemble
coordinates {
(128,575.51)
(256,1005.01)
(512,2212.38)
(1024,4448.00)
};
\legend{$\alpha=1.0$, $\alpha=0.5$}
\end{axis}
\end{tikzpicture}

%% file: figures/fig_elm_ensemble.tex
\begin{tikzpicture}[scale=0.9, every node/.style={scale=1.0}]
    \begin{axis}[
        width  = 0.7*\textwidth,
        height = 10cm,
        ymin=0.0,ymax=1.08,
        ytick={0.0,0.2,0.4,0.6,0.8,1.0},
        major x tick style = transparent,
        ybar=7.0*\pgflinewidth,
        bar width=10.0pt,
%        ymajorgrids = true,
        xlabel = {Number of neurons},
        ylabel = {Testing accuracy},
        symbolic x coords={128,
        				      256,
				      512,
				      1024,
				      2048,
				      4096},
	y tick label style={
    		/pgf/number format/.cd,
   		fixed,
   		fixed zerofill,
    		precision=1},
%	yticklabel pos=right,
        xtick = data,
%        x tick label style={
%        	        rotate=60,
%		font=\footnotesize,
%		anchor=north east,
%		inner sep=0mm
%		},
%		font=\small},
%        scaled y ticks = false,
	%%%%% numbers on bars and rotated
        nodes near coords,
        every node near coord/.append style={/pgf/number format/precision=3,
        								  rotate=90, 
        								  anchor=west,
								  font=\footnotesize},
        %%%%%
        enlarge x limits=0.11,
        legend cell align=left,
        legend style={
%                at={(1,1.05)},
%                anchor=south east,
%	        nodes={rotate=90},%%%%% rotate text in legend
%                at={(0.125,0)},
%                at={(0.125,0)},
%                at={(0.8775,0)},
%                at={(0.135,0.015)},
                at={(0.82,0.015)},
                anchor=south,
                column sep=1ex
        }
    ]
\addplot[fill=blue,opacity=1.00] %%%%% average case
coordinates {
(128,0.697)
(256,0.772)
(512,0.825)
(1024,0.871)
(2048,0.900)
(4096,0.893)
};
\addplot[fill=red,opacity=1.00] %%%%% ensemble
coordinates {
(128,0.784)
(256,0.811)
(512,0.850)
(1024,0.902)
(2048,0.923)
(4096,0.939)
};
\legend{Average case, Ensemble}
\end{axis}
\end{tikzpicture}

%% file: figures/elm2.tex
%\begin{tikzpicture}[scale=0.9]
\begin{tikzpicture}[scale=0.5, every node/.style={scale=0.75}]

    % squares (top)
    \draw[thick,color=blue] (0.5,5.5) rectangle (1.5,6.5);
    \draw[thick,color=blue] (3.0,5.5) rectangle (4.0,6.5);
    \draw[thick,color=blue] (5.5,5.5) rectangle (6.5,6.5);
    \draw[thick,color=blue] (8.0,5.5) rectangle (9.0,6.5);
    \draw[thick,color=blue] (10.5,5.5) rectangle (11.5,6.5);
%    \node at (14.25,6.0) {$\cdots$};
%    \draw[thick,color=blue] (16.25,5.5) rectangle (17.25,6.5);

    \node at (1.0,6.0) {1};
    \node at (3.5,6.0) {2};
    \node at (6.0,6.0) {3};
    \node at (8.5,6.0) {4};
    \node at (11.0,6.0) {5};
%    \node at (16.75,6.0) {$n$};
    
    % squares (middle)
    \draw[thick,color=blue] (3.0,2.5) rectangle (4.0,3.5);
    \draw[thick,color=blue] (5.5,2.5) rectangle (6.5,3.5);
    \draw[thick,color=blue] (8.0,2.5) rectangle (9.0,3.5);
%    \node at (11.0,3.0) {$\cdots$};
%    \draw[thick,color=blue] (13.0,2.5) rectangle (14.0,3.5);

    \node at (3.5,3.0) {1};
    \node at (6.0,3.0) {2};
    \node at (8.5,3.0) {3};
%    \node at (13.5,3.0) {$\ell$};
    
    % squares (bottom)
    \draw[thick,color=blue] (4.0,-0.5) rectangle (5.0,0.5);
    \draw[thick,color=blue] (6.5,-0.5) rectangle (7.5,0.5);
%    \node at (9.5,0.0) {$\cdots$};
%    \draw[thick,color=blue] (12.0,-0.5) rectangle (13.0,0.5);

    \node at (4.5,0.0) {1};
    \node at (7.0,0.0) {2};
%    \node at (12.5,0.0) {$m$};

    % top to middle
    \draw[thick,color=blue,->] (1.0,5.5) -- (3.5,3.5);
    \draw[thick,color=blue,->] (1.0,5.5) -- (6.0,3.5);
    \draw[thick,color=blue,->] (1.0,5.5) -- (8.5,3.5);
%    \draw[thick,color=blue,->] (1.0,5.5) -- (13.5,3.5);

    \draw[thick,color=blue,->] (3.5,5.5) -- (3.5,3.5);
    \draw[thick,color=blue,->] (3.5,5.5) -- (6.0,3.5);
    \draw[thick,color=blue,->] (3.5,5.5) -- (8.5,3.5);
%    \draw[thick,color=blue,->] (3.5,5.5) -- (13.5,3.5);

    \draw[thick,color=blue,->] (6.0,5.5) -- (3.5,3.5);
    \draw[thick,color=blue,->] (6.0,5.5) -- (6.0,3.5);
    \draw[thick,color=blue,->] (6.0,5.5) -- (8.5,3.5);
%    \draw[thick,color=blue,->] (6.0,5.5) -- (13.5,3.5);

    \draw[thick,color=blue,->] (8.5,5.5) -- (3.5,3.5);
    \draw[thick,color=blue,->] (8.5,5.5) -- (6.0,3.5);
    \draw[thick,color=blue,->] (8.5,5.5) -- (8.5,3.5);
%    \draw[thick,color=blue,->] (8.5,5.5) -- (13.5,3.5);

    \draw[thick,color=blue,->] (11.0,5.5) -- (3.5,3.5);
    \draw[thick,color=blue,->] (11.0,5.5) -- (6.0,3.5);
    \draw[thick,color=blue,->] (11.0,5.5) -- (8.5,3.5);
%    \draw[thick,color=blue,->] (11.0,5.5) -- (13.5,3.5);

    % middle to bottom
    \draw[thick,color=blue,->] (3.5,2.5) -- (4.5,0.5);
    \draw[thick,color=blue,->] (3.5,2.5) -- (7.0,0.5);
%    \draw[thick,color=blue,->] (3.5,2.5) -- (12.5,0.5);

    \draw[thick,color=blue,->] (6.0,2.5) -- (4.5,0.5);
    \draw[thick,color=blue,->] (6.0,2.5) -- (7.0,0.5);
%    \draw[thick,color=blue,->] (6.0,2.5) -- (12.5,0.5);

    \draw[thick,color=blue,->] (8.5,2.5) -- (4.5,0.5);
    \draw[thick,color=blue,->] (8.5,2.5) -- (7.0,0.5);
%    \draw[thick,color=blue,->] (8.5,2.5) -- (12.5,0.5);

%    \draw[thick,color=blue,->] (13.5,2.5) -- (4.5,0.5);
%    \draw[thick,color=blue,->] (13.5,2.5) -- (7.0,0.5);
%    \draw[thick,color=blue,->] (13.5,2.5) -- (12.5,0.5);
    
    % labels
%    \node at (18.0,6.0) {$X$};
%    \node at (16.125,4.25) {$W$};
%    \node at (14.75,3.0) {$H$};
%    \node at (13.75,1.5) {$\beta$};
%    \node at (13.75,0.0) {$Y$};
    \node at (-0.1,6.0) {$X$};
    \node at (1.75,4.25) {$W$};
    \node at (2.4,3.0) {$H$};
    \node at (3.25,1.5) {$\beta$};
    \node at (3.4,0.0) {$Y$};

\end{tikzpicture}

%% file: figures/elm2_drop.tex
%\begin{tikzpicture}[scale=0.9]
\begin{tikzpicture}[scale=0.5, every node/.style={scale=0.75}]

    % squares (top)
    \draw[thick,color=blue] (0.5,5.5) rectangle (1.5,6.5);
    \draw[thick,color=blue] (3.0,5.5) rectangle (4.0,6.5);
    \draw[thick,color=blue] (5.5,5.5) rectangle (6.5,6.5);
%    \draw[thick,color=blue] (8.0,5.5) rectangle (9.0,6.5);
    \draw[thick,color=blue,dashed] (8.0,5.5) rectangle (9.0,6.5);
    \draw[thick,color=blue] (10.5,5.5) rectangle (11.5,6.5);
%    \node at (14.25,6.0) {$\cdots$};
%    \draw[thick,color=blue] (16.25,5.5) rectangle (17.25,6.5);

    \node at (1.0,6.0) {1};
    \node at (3.5,6.0) {2};
    \node at (6.0,6.0) {3};
%    \node at (8.5,6.0) {4};
    \node at (8.5,6.0) {\Huge$\times$};
    \node at (11.0,6.0) {5};
%    \node at (16.75,6.0) {$n$};
    
    % squares (middle)
    \draw[thick,color=blue] (3.0,2.5) rectangle (4.0,3.5);
%    \draw[thick,color=blue] (5.5,2.5) rectangle (6.5,3.5);
    \draw[thick,color=blue,dashed] (5.5,2.5) rectangle (6.5,3.5);
    \draw[thick,color=blue] (8.0,2.5) rectangle (9.0,3.5);
%    \node at (11.0,3.0) {$\cdots$};
%    \draw[thick,color=blue] (13.0,2.5) rectangle (14.0,3.5);

    \node at (3.5,3.0) {1};
%    \node at (6.0,3.0) {2};
    \node at (6.0,3.0) {\Huge$\times$};
    \node at (8.5,3.0) {3};
%    \node at (13.5,3.0) {$\ell$};
    
    % squares (bottom)
    \draw[thick,color=blue] (4.0,-0.5) rectangle (5.0,0.5);
    \draw[thick,color=blue] (6.5,-0.5) rectangle (7.5,0.5);
%    \node at (9.5,0.0) {$\cdots$};
%    \draw[thick,color=blue] (12.0,-0.5) rectangle (13.0,0.5);

    \node at (4.5,0.0) {1};
    \node at (7.0,0.0) {2};
%    \node at (12.5,0.0) {$m$};

    % top to middle
    \draw[thick,color=blue,->] (1.0,5.5) -- (3.5,3.5);
%    \draw[thick,color=blue,->] (1.0,5.5) -- (6.0,3.5);
    \draw[thick,color=blue,->] (1.0,5.5) -- (8.5,3.5);
%    \draw[thick,color=blue,->] (1.0,5.5) -- (13.5,3.5);

    \draw[thick,color=blue,->] (3.5,5.5) -- (3.5,3.5);
%    \draw[thick,color=blue,->] (3.5,5.5) -- (6.0,3.5);
    \draw[thick,color=blue,->] (3.5,5.5) -- (8.5,3.5);
%    \draw[thick,color=blue,->] (3.5,5.5) -- (13.5,3.5);

    \draw[thick,color=blue,->] (6.0,5.5) -- (3.5,3.5);
%    \draw[thick,color=blue,->] (6.0,5.5) -- (6.0,3.5);
    \draw[thick,color=blue,->] (6.0,5.5) -- (8.5,3.5);
%    \draw[thick,color=blue,->] (6.0,5.5) -- (13.5,3.5);

%    \draw[thick,color=blue,->] (8.5,5.5) -- (3.5,3.5);
%    \draw[thick,color=blue,->] (8.5,5.5) -- (6.0,3.5);
%    \draw[thick,color=blue,->] (8.5,5.5) -- (8.5,3.5);
%%    \draw[thick,color=blue,->] (8.5,5.5) -- (13.5,3.5);

    \draw[thick,color=blue,->] (11.0,5.5) -- (3.5,3.5);
%    \draw[thick,color=blue,->] (11.0,5.5) -- (6.0,3.5);
    \draw[thick,color=blue,->] (11.0,5.5) -- (8.5,3.5);
%    \draw[thick,color=blue,->] (11.0,5.5) -- (13.5,3.5);

    % middle to bottom
    \draw[thick,color=blue,->] (3.5,2.5) -- (4.5,0.5);
    \draw[thick,color=blue,->] (3.5,2.5) -- (7.0,0.5);
%    \draw[thick,color=blue,->] (3.5,2.5) -- (12.5,0.5);

%    \draw[thick,color=blue,->] (6.0,2.5) -- (4.5,0.5);
%    \draw[thick,color=blue,->] (6.0,2.5) -- (7.0,0.5);
%%    \draw[thick,color=blue,->] (6.0,2.5) -- (12.5,0.5);

    \draw[thick,color=blue,->] (8.5,2.5) -- (4.5,0.5);
    \draw[thick,color=blue,->] (8.5,2.5) -- (7.0,0.5);
%    \draw[thick,color=blue,->] (8.5,2.5) -- (12.5,0.5);

%    \draw[thick,color=blue,->] (13.5,2.5) -- (4.5,0.5);
%    \draw[thick,color=blue,->] (13.5,2.5) -- (7.0,0.5);
%    \draw[thick,color=blue,->] (13.5,2.5) -- (12.5,0.5);
    
    % labels
%    \node at (18.0,6.0) {$X$};
%    \node at (16.125,4.25) {$W$};
%    \node at (14.75,3.0) {$H$};
%    \node at (13.75,1.5) {$\beta$};
%    \node at (13.75,0.0) {$Y$};
    \node at (-0.1,6.0) {$X$};
    \node at (1.75,4.25) {$W$};
    \node at (2.4,3.0) {$H$};
    \node at (3.25,1.5) {$\beta$};
    \node at (3.4,0.0) {$Y$};

\end{tikzpicture}

%% file: figures/fig_elm_dropout.tex
%\begin{tikzpicture}[scale=0.9, every node/.style={scale=1.0}, rotate=-90]
\begin{tikzpicture}[scale=0.9, every node/.style={scale=1.0}]
    \begin{axis}[
        width  = 1.0*\textwidth,
        height = 10cm,
        ymin=0.0,ymax=1.05,
        ytick={0.0,0.2,0.4,0.6,0.8,1.0},
        major x tick style = transparent,
%        ybar=5.0*\pgflinewidth,
        ybar=3.75*\pgflinewidth,
%        ybar,
        bar width=7.0pt,
%        ymajorgrids = true,
        xlabel = {Fold},
        ylabel = {Accuracy},
        symbolic x coords={1,2,3,4,5,6,7,8,9,10},
	y tick label style={
    		/pgf/number format/.cd,
   		fixed,
   		fixed zerofill,
    		precision=1},
%	yticklabel pos=right,
        xtick = data,
%        x tick label style={
%        	        rotate=60,
%        	        rotate=90,
%		font=\footnotesize\tt,
%		anchor=north east,
%		inner sep=0mm
%		},
%		font=\small},
%        scaled y ticks = false,
	%%%%% numbers on bars and rotated
        nodes near coords,
        every node near coord/.append style={/pgf/number format/precision=3,
        								  rotate=90, 
								  scale=0.90,
        								  anchor=west,
								  font=\footnotesize},
        %%%%%
        enlarge x limits=0.07,
        legend cell align=left,
        legend style={
%                at={(1,1.05)},
%                anchor=south east,
%	        nodes={rotate=90},%%%%% rotate text in legend
%                at={(0.125,0)},
%                at={(0.125,0)},
%                at={(0.8775,0)},
%                at={(0.135,0.015)},
                at={(0.805,0.015)},
                anchor=south,
                column sep=1ex
        }
    ]
\addplot[fill=green,opacity=1.00] %%%%% fully connected average
coordinates {
(1,0.872)
(2,0.878)
(3,0.873)
(4,0.873)
(5,0.875)
(6,0.875)
(7,0.872)
(8,0.872)
(9,0.876)
(10,0.877)
};
\addplot[fill=black,opacity=1.00] %%%%% fully connected ensemble
coordinates {
(1,0.900)
(2,0.908)
(3,0.905)
(4,0.899)
(5,0.905)
(6,0.898)
(7,0.895)
(8,0.903)
(9,0.908)
(10,0.906)
};
\addplot[fill=red,opacity=1.00] %%%%% dropout average
coordinates {
(1,0.911)
(2,0.907)
(3,0.907)
(4,0.910)
(5,0.914)
(6,0.911)
(7,0.914)
(8,0.912)
(9,0.914)
(10,0.914)
};
\addplot[fill=blue,opacity=1.00] %%%%% dropout ensemble
coordinates {
(1,0.926)
(2,0.915)
(3,0.923)
(4,0.915)
(5,0.926)
(6,0.922)
(7,0.926)
(8,0.029)
(9,0.925)
(10,0.924)
};
\legend{fully connected average,fully connected ensemble,dropout average,dropout ensemble}
\end{axis}
\end{tikzpicture}

%% file: figures/fig_elm_a.tex
%\begin{tikzpicture}[scale=0.9, every node/.style={scale=1.0}, rotate=-90]
\begin{tikzpicture}[scale=0.85, every node/.style={scale=1.0}]
    \begin{axis}[
        width  = 0.45*\textwidth,
        height = 7cm,
        ymin=0.0,ymax=1.25,
        ytick={0.0,0.2,0.4,0.6,0.8,1.0},
        major x tick style = transparent,
%        ybar=5.0*\pgflinewidth,
%        ybar=3.75*\pgflinewidth,
        ybar,
        bar width=22.5pt,
%        bar width=5.5pt,
%        ymajorgrids = true,
        xlabel = {Number of neurons},
        ylabel = {Accuracy},
        symbolic x coords={512,1024,2048},
	y tick label style={
    		/pgf/number format/.cd,
   		fixed,
   		fixed zerofill,
    		precision=1},
%	yticklabel pos=right,
        xtick = data,
%        x tick label style={
%        	        rotate=60,
%        	        rotate=90,
%		font=\footnotesize\tt,
%		anchor=north east,
%		inner sep=0mm
%		},
%		font=\small},
%        scaled y ticks = false,
	%%%%% numbers on bars and rotated
        nodes near coords,
        every node near coord/.append style={/pgf/number format/precision=4,
        								  rotate=90, 
								  scale=1.0,
        								  anchor=west,
								  font=\footnotesize},
        %%%%%
        enlarge x limits=0.25,
        legend cell align=left,
        legend style={
%                at={(1,1.05)},
%                anchor=south east,
%	        nodes={rotate=90},%%%%% rotate text in legend
%                at={(0.125,0)},
%                at={(0.125,0)},
%                at={(0.8775,0)},
%                at={(0.135,0.015)},
                at={(0.905,0.015)},
                anchor=south,
                column sep=1ex
        }
    ]
\addplot[fill=blue,opacity=1.00]
coordinates {
(512,0.9565)
(1024,0.9662)
(2048,0.9346)
};
%\legend{32 filters,64 filters}
\end{axis}
\end{tikzpicture}

%% file: figures/fig_elm_b.tex
%\begin{tikzpicture}[scale=0.9, every node/.style={scale=1.0}, rotate=-90]
\begin{tikzpicture}[scale=0.85, every node/.style={scale=1.0}]
    \begin{axis}[
        width  = 0.45*\textwidth,
        height = 7cm,
        ymin=0.0,ymax=1.25,
        ytick={0.0,0.2,0.4,0.6,0.8,1.0},
        major x tick style = transparent,
%        ybar=5.0*\pgflinewidth,
%        ybar=3.75*\pgflinewidth,
        ybar,
        bar width=22.5pt,
%        bar width=5.5pt,
%        ymajorgrids = true,
        xlabel = {Input dimension},
        ylabel = {Accuracy},
        symbolic x coords={$512\times 1$,$1024\times 1$,$4096\times 1$},
	y tick label style={
    		/pgf/number format/.cd,
   		fixed,
   		fixed zerofill,
    		precision=1},
%	yticklabel pos=right,
        xtick = data,
%        x tick label style={
%        	        rotate=60,
%        	        rotate=90,
%		font=\footnotesize\tt,
%		anchor=north east,
%		inner sep=0mm
%		},
%		font=\small},
%        scaled y ticks = false,
	%%%%% numbers on bars and rotated
        nodes near coords,
        every node near coord/.append style={/pgf/number format/precision=4,
        								  rotate=90, 
								  scale=1.0,
        								  anchor=west,
								  font=\footnotesize},
        %%%%%
        enlarge x limits=0.25,
        legend cell align=left,
        legend style={
%                at={(1,1.05)},
%                anchor=south east,
%	        nodes={rotate=90},%%%%% rotate text in legend
%                at={(0.125,0)},
%                at={(0.125,0)},
%                at={(0.8775,0)},
%                at={(0.135,0.015)},
                at={(0.905,0.015)},
                anchor=south,
                column sep=1ex
        }
    ]
\addplot[fill=blue,opacity=1.00]
coordinates {
($512\times 1$,0.9662)
($1024\times 1$,0.9647)
($4096\times 1$,0.9650)
};
%\legend{32 filters,64 filters}
\end{axis}
\end{tikzpicture}

%% file: figures/fig_elm_weighted.tex
\begin{tikzpicture}[scale=0.9, every node/.style={scale=1.0}]
    \begin{axis}[
        width  = 0.9*\textwidth,
        height = 10cm,
        ymin=0.0,ymax=1.13,
        ytick={0.0,0.2,0.4,0.6,0.8,1.0},
        major x tick style = transparent,
        ybar=5.0*\pgflinewidth,
        bar width=9.0pt,
%        ymajorgrids = true,
        xlabel = {Fold},
        ylabel = {Accuracy},
        symbolic x coords={1,2,3,4,5,6,7,8,9,10,11,Average},
        xticklabels={1,2,3,4,5,6,7,8,9,10,,Average},
	y tick label style={
    		/pgf/number format/.cd,
   		fixed,
   		fixed zerofill,
    		precision=1},
%	yticklabel pos=right,
        xtick = data,
%        x tick label style={
%        	        rotate=60,
%		font=\footnotesize,
%		anchor=north east,
%		inner sep=0mm
%		},
%		font=\small},
%        scaled y ticks = false,
	%%%%% numbers on bars and rotated
        nodes near coords,
        every node near coord/.append style={/pgf/number format/precision=3,
        								  rotate=90, 
        								  anchor=west,
								  font=\footnotesize},
        %%%%%
        enlarge x limits=0.065,
        legend cell align=left,
        legend style={
%                at={(1,1.05)},
%                anchor=south east,
%	        nodes={rotate=90},%%%%% rotate text in legend
%                at={(0.125,0)},
%                at={(0.125,0)},
%                at={(0.8775,0)},
%                at={(0.135,0.015)},
                at={(0.875,0.015)},
                anchor=south,
                column sep=1ex
        }
    ]
\addplot[fill=blue,opacity=1.00] %%%%% imbalanced
coordinates {
(1,0.963)
(2,0.957)
(3,0.972)
(4,0.963)
(5,0.960)
(6,0.972)
(7,0.959)
(8,0.959)
(9,0.971)
(10,0.971)
(11,0.0)
(Average,0.965)
};
\addplot[fill=red,opacity=1.00] %%%%% balanced
coordinates {
(1,0.975)
(2,0.971)
(3,0.980)
(4,0.979)
(5,0.970)
(6,0.976)
(7,0.979)
(8,0.973)
(9,0.982)
(10,0.981)
(11,0.0)
(Average,0.977)
};
\legend{Imbalanced, Balanced}
\end{axis}
\end{tikzpicture}

%% file: figures/fig_f1.tex
%\begin{tikzpicture}[scale=0.9, every node/.style={scale=1.0}, rotate=-90]
\begin{tikzpicture}[scale=0.9, every node/.style={scale=1.0}]
    \begin{axis}[
        width  = 1.0*\textwidth,
        height = 10cm,
        ymin=0.0,ymax=1.095,
        ytick={0.0,0.2,0.4,0.6,0.8,1.0},
        major x tick style = transparent,
%        ybar=5.0*\pgflinewidth,
        ybar=3.75*\pgflinewidth,
%        ybar,
%        bar width=7.0pt,
        bar width=5.5pt,
%        ymajorgrids = true,
        xlabel = {Malware family},
        ylabel = {F1 score},
        symbolic x coords={Adialer.C,Agent.FYI,Allaple.A,Allaple.L,Alueron.gen!J,Autorun.K,C2LOP.P,C2LOP.gen!g,Dialplatform.B,Dontovo.A,Fakerean,Instantaccess,Lolyda.AA1,Lolyda.AA2,Lolyda.AA3,Lolyda.AT,Malex.gen!J,Obfuscator.AD,Rbot!gen,Skintrim.N,Swizzor.gen!E,Swizzor.gen!I,VB.AT,Wintrim.BX,Yuner.A},
	y tick label style={
    		/pgf/number format/.cd,
   		fixed,
   		fixed zerofill,
    		precision=1},
%	yticklabel pos=right,
        xtick = data,
        x tick label style={
        	        rotate=60,
		font=\footnotesize\tt,
		anchor=north east,
		inner sep=0mm
		},
%		font=\small},
%        scaled y ticks = false,
	%%%%% numbers on bars and rotated
        nodes near coords,
        every node near coord/.append style={/pgf/number format/precision=2,
        								  rotate=90, 
								  scale=0.80,
        								  anchor=west,
								  font=\footnotesize},
        %%%%%
        enlarge x limits=0.025,
        legend cell align=left,
        legend style={
%                at={(1,1.05)},
%                anchor=south east,
%	        nodes={rotate=90},%%%%% rotate text in legend
%                at={(0.125,0)},
%                at={(0.125,0)},
%                at={(0.8775,0)},
%                at={(0.135,0.015)},
                at={(0.93,0.015)},
                anchor=south,
                column sep=1ex
        }
    ]
\addplot[fill=blue,opacity=1.00] %%%%% 32 filters
coordinates {
(Adialer.C,1.00)
(Agent.FYI,1.00)
(Allaple.A,1.00)
(Allaple.L,1.00)
(Alueron.gen!J,0.94)
(Autorun.K,0.00)
(C2LOP.P,0.76)
(C2LOP.gen!g,0.75)
(Dialplatform.B,1.00)
(Dontovo.A,1.00)
(Fakerean,1.00)
(Instantaccess,1.00)
(Lolyda.AA1,1.00)
(Lolyda.AA2,0.97)
(Lolyda.AA3,0.90)
(Lolyda.AT,0.96)
(Malex.gen!J,0.96)
(Obfuscator.AD,1.00)
(Rbot!gen,0.87)
(Skintrim.N,1.00)
(Swizzor.gen!E,0.67)
(Swizzor.gen!I,0.61)
(VB.AT,0.98)
(Wintrim.BX,0.93)
(Yuner.A,0.95)
};
\addplot[fill=red,opacity=1.00] %%%%% 64 filters
coordinates {
(Adialer.C,1.00)
(Agent.FYI,0.96)
(Allaple.A,0.99)
(Allaple.L,0.99)
(Alueron.gen!J,1.00)
(Autorun.K,1.00)
(C2LOP.P,0.90)
(C2LOP.gen!g,0.90)
(Dialplatform.B,1.00)
(Dontovo.A,1.00)
(Fakerean,1.00)
(Instantaccess,1.00)
(Lolyda.AA1,1.00)
(Lolyda.AA2,1.00)
(Lolyda.AA3,1.00)
(Lolyda.AT,1.00)
(Malex.gen!J,0.90)
(Obfuscator.AD,1.00)
(Rbot!gen,1.00)
(Skintrim.N,1.00)
(Swizzor.gen!E,0.70)
(Swizzor.gen!I,0.50)
(VB.AT,1.00)
(Wintrim.BX,1.00)
(Yuner.A,1.00)
};
\legend{CNN, ELM}
\end{axis}
\end{tikzpicture}

%% file: figures/fig_cnn_elm_time2.tex
\begin{tikzpicture}[scale=0.9, every node/.style={scale=1.0}]
    \begin{axis}[
        width = 0.6*\textwidth,
        height = 8cm,
        ymin=0.0,ymax=415,
        ytick={0,50,100,150,200,250,300,350},
        major x tick style = transparent,
        ybar=7.0*\pgflinewidth,
%	ybar=0pt,
%    	bar shift=0pt,
	bar width=28.0pt,
%        ymajorgrids = true,
        xlabel = {Architecture},
        ylabel = {Time (minutes)},
%        symbolic x coords={CNN-2,CNN-1,ELM-0,ELM-1},
	y tick label style={
    		/pgf/number format/.cd,
		1000 sep={},
   		fixed,
   		fixed zerofill,
    		precision=0},
%	yticklabel pos=right,
%        xtick = data,
%	xtick={CNN-2,CNN-1,ELM-0,ELM-1},
	xtick={1,2,3,4},
	xticklabels={},
%        x tick label style={
%        	        rotate=60,
%		font=\footnotesize,
%		anchor=north east,
%		inner sep=0mm
%		},
%		font=\small},
%        scaled y ticks = false,
	%%%%% numbers on bars and rotated
        nodes near coords,
        every node near coord/.append style={/pgf/number format/precision=2,
        								  rotate=90, 
        								  anchor=west,
								  font=\footnotesize,
								  /pgf/number format/1000 sep=},
        %%%%%
        enlarge x limits=0.9,
        legend cell align=left,
        legend style={
%                at={(1,1.05)},
%                anchor=south east,
%	        nodes={rotate=90},%%%%% rotate text in legend
%                at={(0.125,0)},
%                at={(0.125,0)},
                at={(0.76,0.635)},
                anchor=south,
                column sep=1ex
        }
    ]
\addplot[fill=yellow,opacity=1.00,xticklabels={CNN}] %%%%% 
coordinates {
%(CNN-2,334.50)% 2-layer CNN
(1,334.50)% 2-layer CNN
};
\addlegendentry{CNN (2 layers)},
\addplot[fill=green,opacity=1.00] %%%%% 
coordinates {
%(CNN-1,254.75)% 1-layer CNN
(2,254.75)% 1-layer CNN
};
\addlegendentry{CNN (1 layer)},
\addplot[fill=red,opacity=1.00] %%%%% 
coordinates {
%(ELM-0,74.12)% ELM alpha = 0
(3,74.12)% ELM alpha = 0
};
\addlegendentry{ELM ($\alpha=0$)}
\addplot[fill=blue,opacity=1.00] %%%%% 
coordinates {
%(ELM-1,4.91)% ELM alpha = 1
(4,4.91)% ELM alpha = 1
};
\addlegendentry{ELM ($\alpha=1$)}
\end{axis}
\end{tikzpicture}